\documentclass[pdftex,twocolumn,epjc3]{svjour3}
\pdfoutput=1
\RequirePackage[T1]{fontenc}
\smartqed
\RequirePackage{amsmath}
\RequirePackage{graphicx}
\RequirePackage{mathptmx}
\RequirePackage[numbers,sort&compress]{natbib}

\usepackage[hyperindex,breaklinks]{hyperref}
\usepackage{preprintcover}
\PreprintCoverPaperTitle{Measurement of jet shapes in top pair events at $\boldsymbol{\sqrt{s} = 7 \TeV}$ using the ATLAS detector}
\PreprintIdNumber{CERN-PH-EP-2013-067}
\PreprintCoverAbstract{A measurement of jet shapes in top-quark pair events using $1.8 \ \ifb$ of $\sqrt{s} = 7 \TeV$ $pp$ collision data recorded by the ATLAS detector at the LHC is presented. Samples of top-quark pair events are selected in both the single-lepton and dilepton final states. The differential and integrated shapes of the jets initiated by bottom-quarks from the top-quark decays are compared with those of the jets originated by light-quarks from the hadronic $\Wboson$-boson decays $\Wboson\rightarrow q\bar{q'}$ in the single-lepton channel. The light-quark jets are found to have a narrower distribution of the momentum flow inside the jet area than $b$-quark jets.}

\PreprintJournalName{Eur. Phys. J. C}

\usepackage{verbatim}
\usepackage{float}
\usepackage{atlasphysics}
\begin{document}

\title{Measurement of jet shapes in top-quark pair events at $\boldsymbol{\sqrt{s} = 7 \TeV}$ using the ATLAS detector}

\author{The ATLAS Collaboration\thanksref{addr1}}
\institute{CERN, 1211 Geneva 23, Switzerland\label{addr1}}

\date{Received: date / Accepted: date}
\maketitle

\begin{abstract}
A measurement of jet shapes in top-quark pair events using $1.8 \ \ifb$ of $\sqrt{s} = 7 \TeV$ $pp$ collision data recorded by the ATLAS detector at the LHC is presented. Samples of top-quark pair events are selected in both the single-lepton and dilepton final states. The differential and integrated shapes of the jets initiated by bottom-quarks from the top-quark decays are compared with those of the jets originated by light-quarks from the hadronic $\Wboson$-boson decays $\Wboson\rightarrow q\bar{q'}$ in the single-lepton channel. The light-quark jets are found to have a narrower distribution of the momentum flow inside the jet area than $b$-quark jets.
\end{abstract}
\section{Introduction}
\label{sec1}
Hadronic jets are observed in large momentum-transfer interactions. They are theoretically interpreted to arise when partons -- quarks ($q$) and gluons ($g$) -- are emitted in collision events of subatomic particles. Partons then evolve into hadronic jets in a two-step process. The first can be described by perturbation theory and gives rise to a parton shower, the second is non-perturbative and is responsible for the hadronisation. The internal structure of a jet is expected to depend primarily on the type of parton it originated from, with some residual dependence on the quark production and fragmentation process. For instance, due to the different colour factors in $ggg$ and $qqg$ vertices, gluons lead to more parton radiation and therefore gluon-initiated jets are expected to be broader than quark-initiated jets.\\
For jets defined using cone or $k_t$ algorithms \cite{cone, kt}, jet shapes, i.e. the normalised transverse momentum flow as a function of the distance to the jet axis \cite{ellis}, have been traditionally used as a means of understanding the evolution of partons into hadrons in $e^+ e^-$, $ep$ and hadron colliders \cite{cdf01,d0,lep,hera2,h1,chek,lhc,cms}. It is experimentally observed that jets in $e^+ e^-$ and $ep$ are narrower than those observed in $p\bar{p}$ and $pp$ collisions and this is interpreted as a result of the different admixtures of quark and gluon jets present in these different types of interactions \cite{kellis}. Furthermore, at high momentum transfer, where fragmentation effects are less relevant, jet shapes have been found to be in qualitative agreement with next-to-leading-order (NLO) QCD predictions and in quantitative agreement with those including leading logarithm corrections \cite{leadLog}. Jet shapes have also been proposed as a tool for studies of substructure or in searches for new phenomena in final states with highly boosted particles \cite{boost1,boost2,boost3,gluinos}.\\
Due to the mass of the $b$-quark, jets originating from a $b$-quark (hereafter called $b$-jets) are expected to be broader than light-quark jets, including charm jets, hereafter called light jets. This expectation is supported by observations by the CDF collaboration in Ref. \cite{CDF}, where a comparison is presented between jet shapes in a $b$-jet enriched sample with a purity of roughly 25\% and an inclusive sample where no distinction is made between the flavours.\\
This paper presents the first measurement of $b$-jet shapes in top pair events. The $t\bar{t}$ final states are a source of $b$-jets, as the top quark decays almost exclusively via $t\rightarrow \Wboson b$. While the dilepton channel, where both $\Wboson$ bosons decay to leptons, is a very pure source of $b$-jets, the single-lepton channel contains $b$-jets and light jets, the latter originating from the dominant $\Wboson^+ \rightarrow u\bar{d}, c\bar{s}$ decays and their charge conjugates.
A comparison of the light- and $b$-jet shapes measured in the $t\bar{t}$ decays improves the CDF measurement discussed above, as the jet purity achieved using $t\bar{t}$ events is much higher. In addition, these measurements could be used to improve the modelling of jets in $t\bar{t}$ production Monte Carlo (MC) models in a new kinematic regime.\\
This paper is organised as follows. In Sect. \ref{sec2} the ATLAS detector is described, while Sect. \ref{sec3} is dedicated to the MC samples used in the analysis. In Sects. \ref{sec4} and \ref{sec5}, the physics object and event selection for both the dilepton and single-lepton $t\bar{t}$ samples is presented. Section \ref{sec6} is devoted to the description of both the $b$-jet and light-jet samples obtained in the single-lepton final state. The differential and the integrated shape distributions of these jets are derived in Sect. \ref{sec7}. In Sect. \ref{sec8} the results on the average values of the jet shape variables at the detector level are presented, including those for the $b$-jets in the dilepton channel. Results corrected for detector effects are presented in Sect. \ref{sec9}. In Sect. \ref{sec10} the systematic uncertainties are discussed, and Sect. \ref{sec11} contains a discussion of the results. Finally, Sect. \ref{sec12} includes the summary and conclusions.
\section{The ATLAS detector}
\label{sec2}
The ATLAS detector \cite{detector} is a multi-purpose particle physics detector with a forward-backward symmetric cylindrical geometry \footnote{ATLAS uses a right-handed coordinate system with its origin at the nominal interaction point (IP) in the centre of the detector and the $z$-axis along the beam pipe. The $x$-axis points from the IP to the centre of the LHC ring, and the $y$-axis points upward. Cylindrical coordinates $(r,\phi)$ are used in the transverse plane, $\phi$ being the azimuthal angle around the beam pipe. The pseudorapidity is defined in terms of the polar angle $\theta$ as $\eta=-\ln\tan(\theta/2)$.} and a solid angle coverage of almost $4\pi$.\\
The inner tracking system covers the pseudorapidity range $|\eta|< 2.5$, and consists of a silicon pixel detector, a silicon microstrip detector, and, for $|\eta|<2.0$, a transition radiation tracker. The inner detector (ID) is surrounded by a thin superconducting solenoid providing a 2 $\mathrm{T}$ magnetic field along the beam direction. A high-granularity liquid-argon sampling electromagnetic calorimeter covers the region $|\eta|<3.2$. An iron/scintillator tile hadronic calorimeter provides coverage in the range $|\eta|<1.7$. The endcap and forward regions, spanning $1.5<|\eta|<4.9$, are instrumented with liquid-argon calorimeters for electromagnetic and hadronic measurements. The muon spectrometer surrounds the calorimeters. It consists of three large air-core superconducting toroid systems and separate trigger and high-precision tracking chambers providing accurate muon tracking for $|\eta|<2.7$.\\
The trigger system \cite{atlasTrigger} has three consecutive levels: level 1 (L1), level 2 (L2) and the event filter (EF). The L1 triggers are hardware-based and use coarse detector information to identify regions of interest, whereas the L2 triggers are based on fast software-based online data reconstruction algorithms. Finally, the EF triggers use offline data reconstruction algorithms. For this analysis, the relevant triggers select events with at least one electron or muon.
\section{Monte Carlo Samples}
\label{sec3}
Monte Carlo generators are used in which $t\bar{t}$ production is implemented with matrix elements calculated up to NLO accuracy. The generated events are then passed through a detailed \textsc{Geant4} simulation \cite{geant1,geant2} of the ATLAS detector.
The baseline MC samples used here are produced with the \textsc{MC@NLO} \cite{mcnlo} or \textsc{Powheg} \cite{powheg} generators for the matrix element calculation; the parton shower and hadronisation processes are implemented with \textsc{Herwig} \cite{Herwig} using the cluster hadronisation model \cite{cluster} and \textsc{CTEQ6.6} \cite{pdf1} parton distribution functions (PDFs). Multi-parton interactions are simulated using \textsc{Jimmy} \cite{Jimmy} with the AUET1 tune \cite{auet1}. This MC generator package has been used for the description of the $t\bar{t}$ final states for ATLAS measurements of the cross section \cite{xsec1l,xsec2l} and studies of the kinematics \cite{spinCorr}. \\
Additional MC samples are used to check the hadronisation model dependence of the jet shapes. They are based on \textsc{Powheg+Pythia} \cite{powheg,pythia}, with the \textsc{MRST2007LO*} PDFs \cite{pdf2}. The \textsc{AcerMC} generator \cite{acer} interfaced to \textsc{Pythia} with the \textsc{Perugia 2010} tune \cite{Perugia} for parton showering and hadronisation is also used for comparison. Here the parton showers are ordered by transverse momentum and the hadronisation proceeds through the Lund string fragmentation scheme \cite{lund}. The underlying event and other soft effects are simulated by \textsc{Pythia} with the AMBT1 tune \cite{ambt1}. Comparisons of different event generators show that jet shapes in top-quark decays show little sensitivity to initial-state radiation effects, different PDF choices or underlying-event effects. They are more sensitive to details of the parton shower and the fragmentation scheme.\\
Samples of events including $\Wboson$ and $\Zboson$ bosons produced in association with light- and heavy-flavour jets are generated using the \textsc{Alpgen} \cite{alpgen} generator with the \textsc{CTEQ6L} PDFs \cite{pdf3}, and interfaced with \textsc{Herwig} and \textsc{Jimmy}. The same generator is used for the diboson backgrounds, $\Wboson\Wboson$, $\Wboson\Zboson$ and $\Zboson\Zboson$, while \textsc{MC@NLO} is used for the simulation of the single-top backgrounds, including the $\mathrm{t}$- and $\mathrm{s}$-channels as well as the $\Wboson t$-channel.\\
The MC-simulated samples are normalised to the corresponding cross sections. The $t\bar{t}$ signal is normalised to the cross section calculated at approximate next-to-next-to-leading order (NNLO) using the \textsc{Hathor} package \cite{HATHOR}, while for the single-top production cross section, the calculations in Refs. \cite{kidonakis1,kidonakis2,kidonakis3} are used. The $\Wboson$+jets and $\Zboson$+jets cross sections are taken from \textsc{Alpgen} \cite{alpgen} with additional NNLO $K$-factors as given in Ref. \cite{fewz}.\\
The simulated events are weighted such that the distribution of the number of interactions per bunch crossing in the simulated samples matches that of the data. Finally, additional correction factors are applied to take into account the different object efficiencies in data and simulation. The scale factors used for these corrections tipically differ from unity by 1\% for electrons and muons, and by a few percent for $b$-tagging.
\section{Physics object selection}
\label{sec4}
Electron candidates are reconstructed from energy deposits in the calorimeter that are associated with tracks reconstructed in the ID. The candidates must pass a tight selection \cite{elec}, which uses calorimeter and tracking variables as well as transition radiation for $|\eta| < 2.0$, and are required to have transverse momentum $p_{\mathrm{T}} > 25 \GeV$ and $|\eta| < 2.47$. Electrons in the transition region between the barrel and endcap calorimeters, $1.37 < |\eta| < 1.52$, are not considered.\\
Muon candidates are reconstructed by searching for track segments in different layers of the muon spectrometer. These segments are combined and matched with tracks found in the ID. The candidates are refitted using the complete track information from both detector systems and are required to have a good fit and to satisfy $p_{\mathrm{T}} > 20 \GeV$ and $|\eta| < 2.5$.\\
Electron and muon candidates are required to be isolated to reduce backgrounds arising from jets and to suppress the selection of leptons from heavy-flavour semileptonic decays. For electron candidates, the transverse energy deposited in the calorimeter and which is not associated with the electron itself ($E^{\mathrm{iso}}_{\mathrm{T}}$) is summed in a cone in $\eta-\phi$ space of radius \footnote{The radius in the $\eta-\phi$ space is defined as $\Delta R = \sqrt{(\Delta\eta)^2+(\Delta\phi)^2}$} $\Delta R = 0.2$ around the electron. The $E^{\mathrm{iso}}_{\mathrm{T}}$ value is required to be less than $3.5 \GeV$. For muon candidates, both the corresponding calorimeter isolation $E^{\mathrm{iso}}_{\mathrm{T}}$ and the analogous track isolation transverse momentum ($p^{\mathrm{iso}}_{\mathrm{T}}$) must be less than $4 \GeV$ in a cone of $\Delta R = 0.3$. The track isolation is calculated from the scalar sum of the transverse momenta of tracks with $p_{\mathrm{T}} > 1 \GeV$, excluding the muon.\\
Muon candidates arising from cosmic rays are rejected by removing candidate pairs that are back-to-back in the transverse plane and that have transverse impact parameter relative to the beam axis $|d_0| > 0.5$ mm.\\
Jets are reconstructed with the anti-$k_{t}$ algorithm \cite{jets,fastjet} with radius parameter $R = 0.4$. 
This choice for the radius has been used in measurements of the top-quark mass \cite{topmass} and also in multi-jet cross-section measurements \cite{atlasJets}. The inputs to the jet algorithm are topological clusters of calorimeter cells. These clusters are seeded by calorimeter cells with energy $|E_{\mathrm{cell}}| > 4 \sigma$ , where $\sigma$ is the cell-by-cell RMS of the noise (electronics plus pileup). Neighbouring cells are added if $|E_{\mathrm{cell}}| > 2 \sigma$ and clusters are formed through an iterative procedure \cite{lampl}. In a final step, all remaining neighbouring cells are added to the cluster.\\
The baseline calibration for these clusters calculates their energy using the electromagnetic energy scale \cite{JES}. This is established using test-beam measurements for electrons and muons in the electromagnetic and hadronic calorimeters \cite{lampl,aleksa,aharouche}. Effects due to the differing response to electromagnetic and hadronic showers, energy losses in the dead material, shower leakage, as well as inefficiencies in energy clustering and jet reconstruction are also taken into account. This is done by matching calorimeter jets with MC particle jets in bins of $\eta$ and $E$, and supplemented by in situ calibration methods such as jet momentum imbalance in $\Zboson / \gamma^{*}$ + 1 jet events. This is called the Jet Energy Scale (JES) calibration, thoroughly discussed in Ref. \cite{JES}. The JES uncertainty contains an extra term for $b$-quark jets, as the jet response is different for $b$-jets and light jets because they have different particle composition. References \cite{atlasJets} and \cite{atlasBxsec} contain more details on the JES and a discussion of its uncertainties.\\
Jets that overlap with a selected electron are removed if they are closer than $\Delta R = 0.2$, while if a jet is closer than $\Delta R = 0.4$ to a muon, the muon is removed.\\
The primary vertex is defined as the $pp$ interaction vertex with the largest $\sum_{i} p_{\mathrm{T}i}^2$, where the sum runs over the tracks with $\pt > 150 \MeV$ associated with the vertex.\\
Jets are identified as candidates for having originated from a $b$-quark ($b$-tagged) by an algorithm based on a neural-network approach, as discussed in Sect. \ref{sec6}.\\
The reconstruction of the direction and magnitude ($\met{}$) of the missing transverse momentum is described in Ref. \cite{etmiss} and begins with the vector sum of the transverse momenta of all jets with $p_{\mathrm{T}} > 20 \GeV$ and $|\eta| < 4.5$. The transverse momenta of electron candidates are added. The contributions from all muon candidates and from all calorimeter clusters not belonging to a reconstructed object are also included.
\section{Event selection}
\label{sec5}
Two samples of events are selected: a dilepton sample, where both $\Wboson$ bosons decay to leptons ($e$, $\mu$, including leptonic $\tau$ decays), and a single-lepton sample, where one $\Wboson$ boson decays to leptons and the other to a $q\bar{q'}$ pair, giving rise to two more jets (see Fig. \ref{fig:feyn}).
The selection criteria follow those in Ref. \cite{xsec1l} for the single-lepton sample and Ref. \cite{xsec2l} for the dilepton sample. Events are triggered by inclusive high-$p_{\mathrm{T}}$ electron or muon EF triggers. The trigger thresholds are $18\GeV$ for muons and $20\GeV$ for electrons. The dataset used for the analysis corresponds to the first half of the data collected in 2011, with a centre-of-mass energy $\sqrt{s} = 7 \TeV$ and an integrated luminosity of $1.8\ \ifb$. This data-taking period is characterised by an instantaneous luminosity smaller than $1.5\times 10^{33}$ cm$^{-2}$ s$^{-1}$, for which the mean number of interactions per bunch crossing is less than six. To reject the non-collision background, the primary vertex is required to have at least four tracks, each with $p_{\mathrm{T}} > 150 \MeV$, associated with it. Pile-up effects are therefore small and have been taken into account as a systematic uncertainty.
\begin{figure*}
\centering
\includegraphics[width=6.8cm,height=4.15cm]{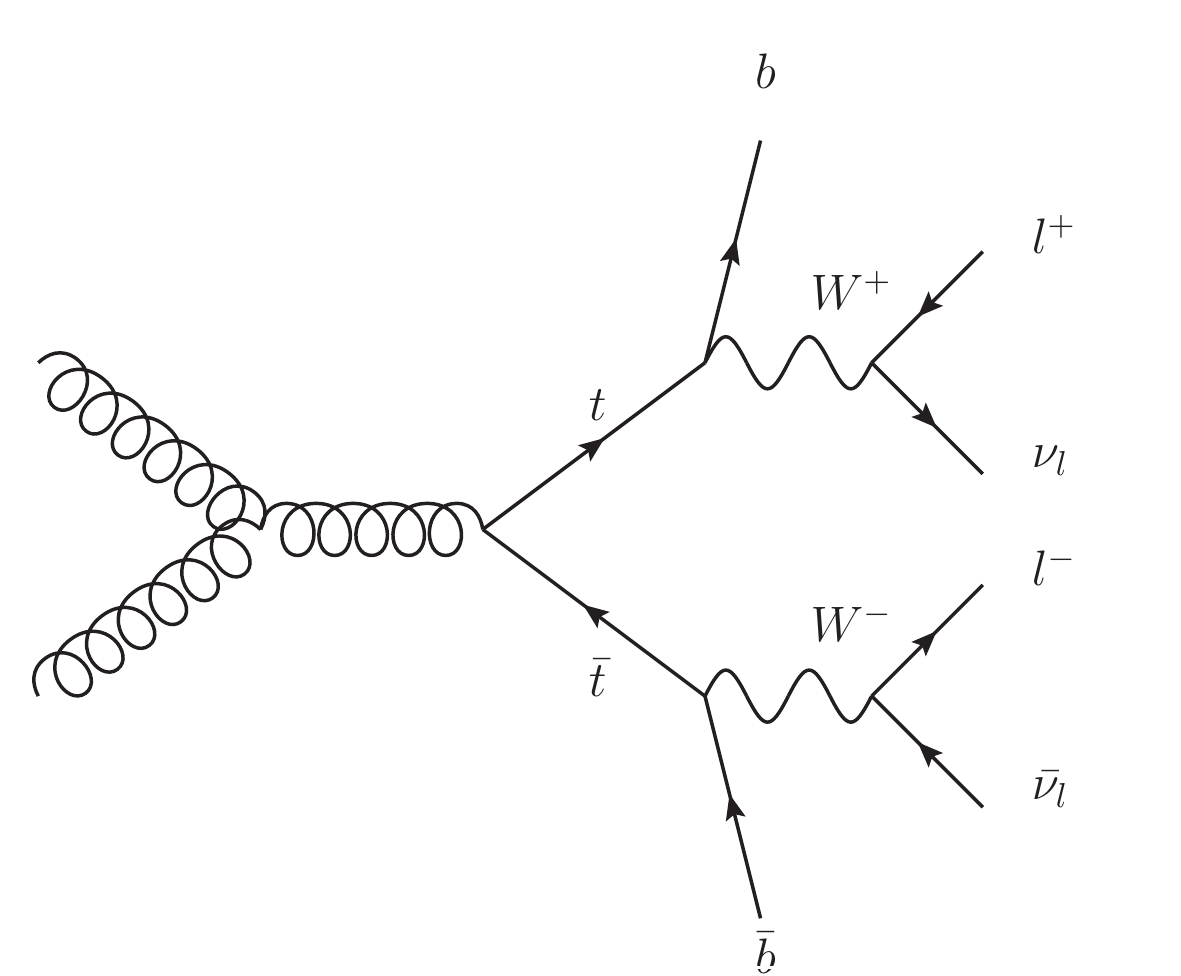}
\hspace{0.8cm}
\includegraphics[width=6.8cm,height=4.15cm]{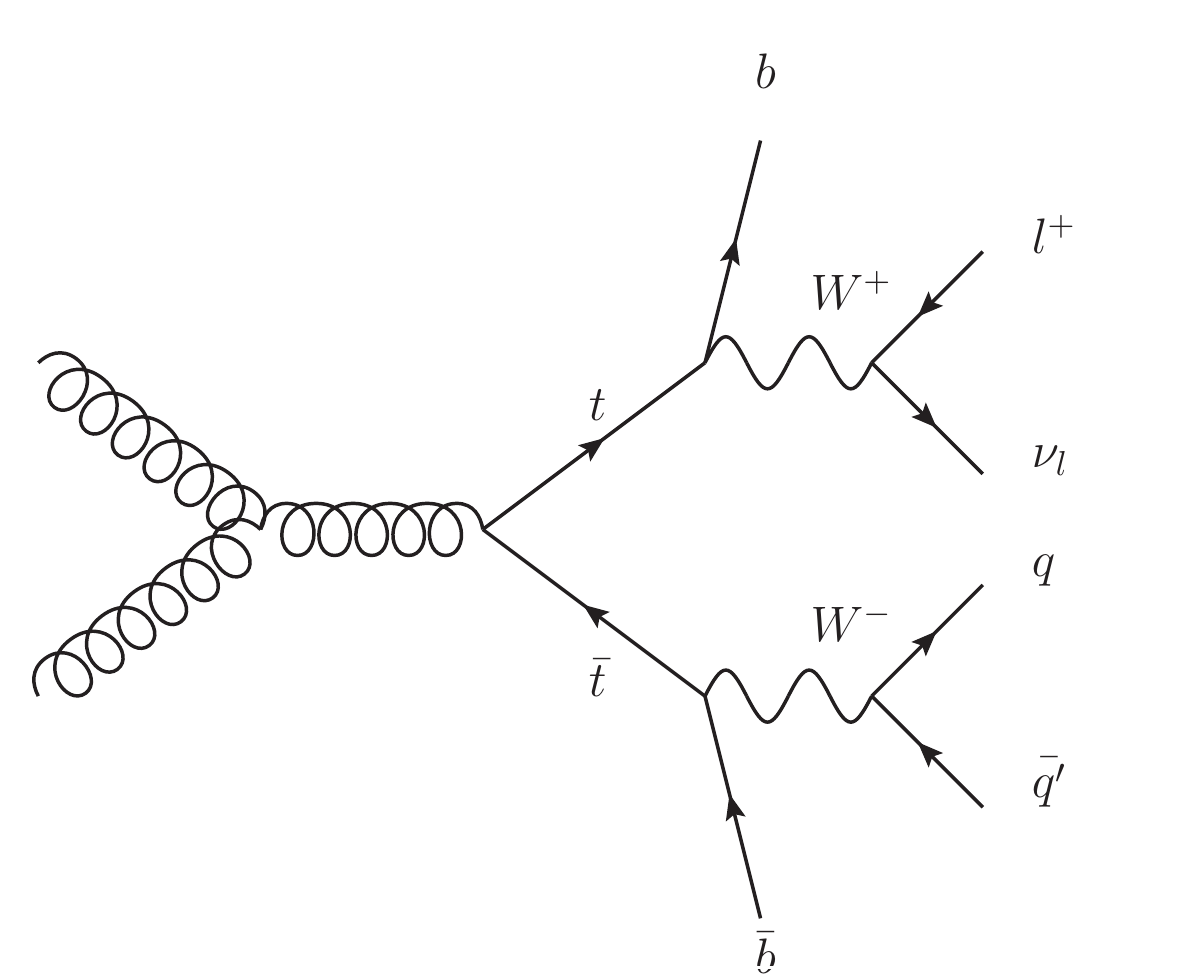}
\caption{Example LO Feynman diagrams for $gg\rightarrow t\bar{t}$ in the dilepton (left) and single-lepton (right) decay modes.}
\label{fig:feyn}
\end{figure*}
\subsection{Dilepton sample}
In the dilepton sample, events are required to have two charged leptons and $\met{}$ from the leptonic $\Wboson$-boson decays to a neutrino and an electron or muon. The offline lepton selection requires two isolated leptons ($e$ or $\mu$) with opposite charge and with transverse momenta $p_{\mathrm{T}}(e) > 25 \GeV$, where $p_{\mathrm{T}}(e) = E_{\mbox{\small{cluster}}}\sin(\theta_{\mbox{\small{track}}})$, $E_{\mbox{\small{cluster}}}$ being the cluster energy and $\theta_{\mbox{\small{track}}}$ the track polar angle, and  $p_{\mathrm{T}}(\mu) > 20 \GeV$. At least one of the selected leptons has to match the corresponding trigger object.\\
Events are further filtered by requiring at least two jets with $\pt > 25 \GeV$ and $\left|\eta\right| < 2.5$ in the event. In addition, at least one of the selected jets has to be tagged as a $b$-jet, as discussed in the next section. The whole event is rejected if a jet is identified as an out-of-time signal or as noise in the calorimeter.\\
The missing transverse momentum requirement is $\met{} > 60 \GeV$ for the $ee$ and $\mu\mu$ channels. For the $e\mu$ channel, $H_{\mathrm{T}}$ is required to be greater than 130 \GeV, where $H_{\mathrm{T}}$ is the scalar sum of the $p_{\mathrm{T}}$ of all muons, electrons and jets.
To reject the Drell--Yan lepton pair background in the $ee$ and $\mu\mu$ channels, the lepton pair is required to have an invariant mass $m_{\ell\ell}$ greater than 15 $\GeV$ and to lie outside of a $\Zboson$-boson mass window, rejecting all events where the two-lepton invariant mass satisfies $\left|m_{\ell\ell}-m_Z\right| < 10 \GeV$.\\
The selected sample consists of 95\% $t\bar{t}$ events, but also backgrounds from the final states $\Wboson$ + jets and $\Zboson$ + jets, where the gauge bosons decay to leptons. All backgrounds, with the exception of multi-jet production, have been estimated using MC samples. The multi-jet background has been estimated using the jet--electron method \cite{qcdBkg}. This method relies on the identification of jets which, due to their high electromagnetic energy fraction, can fake electron candidates. The jet--electron method is applied with some modifications to the muon channel as well. The normalisation is estimated using a binned likelihood fit to the $\met{}$ distribution. The results are summarised in Table \ref{table1}.
\begin{table}
\caption{The expected composition of the dilepton sample. Fractions are relative to the total number of expected events. `Other EW' corresponds to the $\Wboson$ + jets and diboson ($\Wboson\Wboson$, $\Wboson\Zboson$ and $\Zboson\Zboson$) contributions.}
\label{table1}
\begin{center}
\begin{tabular}{ccc}
\hline
Process & Expected events & Fraction\\
\hline
$t\bar{t}$ &  2100 $\pm$ 110 & 94.9\%\\
$\Zboson$ + jets ($\Zboson \rightarrow \ell^{+}\ell^{-}$) & 14 $\pm$ 1 & 0.6\%\\
Other EW (\Wboson, diboson) & 4 $\pm$ 2 & 0.2\%\\
Single top & 95 $\pm$ 2 & 4.3\%\\
Multi-jet & $0^{+2}_{-0}$ & 0.0\%\\
\hline
\bf{Total Expected} & 2210 $\pm$ 110 & \\
\hline
\bf{Total Observed} & 2067 & \\
\end{tabular}
\end{center}
\end{table}
\subsection{Single-Lepton sample}
In this case, the event is required to have exactly one isolated lepton with $p_{\mathrm{T}} > 25 \GeV$ for electrons and $p_{\mathrm{T}} > 20 \GeV$ for muons.
To account for the neutrino in the leptonic $\Wboson$ decay, $\met{}$ is required to be greater than  $35 \GeV$ in the electron channel and greater than $20 \GeV$ in the muon channel. The $\met{}$ resolution is below 10 $\GeV$ \cite{etmiss}. Furthermore, the transverse mass \footnote{The transverse mass is defined as $m_{\mathrm{T}} = \sqrt{2p_{\mathrm{T}}^\ell \met{}(1-\cos\Delta\phi_{\ell\nu})}$, where $\Delta\phi_{\ell\nu}$ is the angle in the transverse plane between the selected lepton and the $\met{}$ direction.} ($m_{\mathrm{T}}$) is required to be greater than $25 \GeV$ in the $e$-channel and to satisfy the condition $\met{}+m_{\mathrm{T}} > 60 \GeV$ in the $\mu$-channel.\\
The jet selection requires at least four jets ($p_{\mathrm{T}} > 25 \GeV$ and $\left|\eta\right| < 2.5$) in the final state, and at least one of them has to be tagged as a $b$-jet. The fraction of $t\bar{t}$ events in the sample is 77\%; the main background contributions for the single-lepton channel have been studied as in the previous case, and are summarised in Table \ref{table2}. As in the dileptonic case, the multi-jet background has been estimated using the jet--electron method.
\begin{table}
\caption{The expected composition of the single-lepton sample. Fractions are relative to the total number of expected events. In this case `Other EW' includes $\Zboson$ + jets and diboson processes.}
\label{table2}
\begin{center}
\begin{tabular}{ccc}
\hline
Process & Expected events & Fraction\\
\hline
$t\bar{t}$ & 14000 $\pm$ 700 & 77.4\%\\
$\Wboson$ + jets ($\Wboson\rightarrow \ell\nu$) & 2310 $\pm$ 280 & 12.8\%\\
Other EW (\Zboson, diboson) & 198 $\pm$ 18 & 1.1\%\\
Single top & 668 $\pm$ 14 & 3.7\%\\
Multi-jet & 900 $\pm$ 450 & 5.0\%\\
\hline
\bf{Total Expected} & 18000 $\pm$ 900 &\\
\hline
\bf{Total Observed} & 17019 & \\
\end{tabular}
\end{center}
\end{table}
\section{Jet sample definition}
\label{sec6}
Jets reconstructed in the single-lepton and dilepton samples are now subdivided into $b$-jet and light-jet samples. In order to avoid contributions from non-primary collisions, it is required that the jet vertex fraction (JVF) be greater than 0.75. After summing the scalar $p_{\mathrm{T}}$ of all tracks in a jet, the JVF is defined as the fraction of the total scalar $p_{\mathrm{T}}$ that belongs to tracks originating from the primary vertex. This makes the average jet multiplicity independent of the number of $pp$ interaction vertices. This selection is not applied to jets with no associated tracks. Also, to reduce the impact of pileup on the jets, the $p_{\mathrm{T}}$ threshold has been raised to $30 \GeV$.\\
Jets whose axes are closer than $\Delta R = 0.8$, which is twice the jet radius, to some other jet in the event are not considered. This is done to avoid possible overlaps between the jet cones, which would bias the shape measurement. These configurations are typical in boosted $\Wboson$ bosons, leading to light jets which are not well separated. The resulting $\Delta R$ distributions for any pair of $b$-jets or light jets are approximately constant between 0.8 and $\pi$ and exhibit an exponential fall-off between $\pi$ and the endpoint of the distribution.
\subsection{$b$-jet samples}
To select $b$-jets, a neural-network algorithm, which relies on the reconstruction of secondary vertices and impact parameter information in the three spatial dimensions, is used. The reconstruction of the secondary decay vertices makes use of an iterative Kalman-filter algorithm \cite{kalman} which relies on the hypothesis that the $b\to c\to X$ decay chains lie in a straight line originally taken to be parallel to the jet axis. The working point of the algorithm is chosen to maximise the purity of the sample. It corresponds to a $b$-tagging efficiency of 57\% for jets originating from $b$-quarks in simulated $t\bar{t}$ events, and a $u,d,s$-quark jet rejection factor of about 400, as well as a $c$-jet rejection factor of about 10 \cite{btagAtl,btagAtl2}.
The resulting number of $b$-jets selected in the dilepton (single-lepton) sample is 2279 (16735). A second working point with a $b$-tagging efficiency of 70\% is also used in order to evaluate the dependence of the measured jet shapes on $b$-tagging.\\
Figure \ref{fig:bsample_ds} shows the $b$-tagged jet transverse momentum distributions for the single-lepton and dilepton channels.
\begin{figure}
\centering
\includegraphics[width=8.5cm,height=6.0cm]{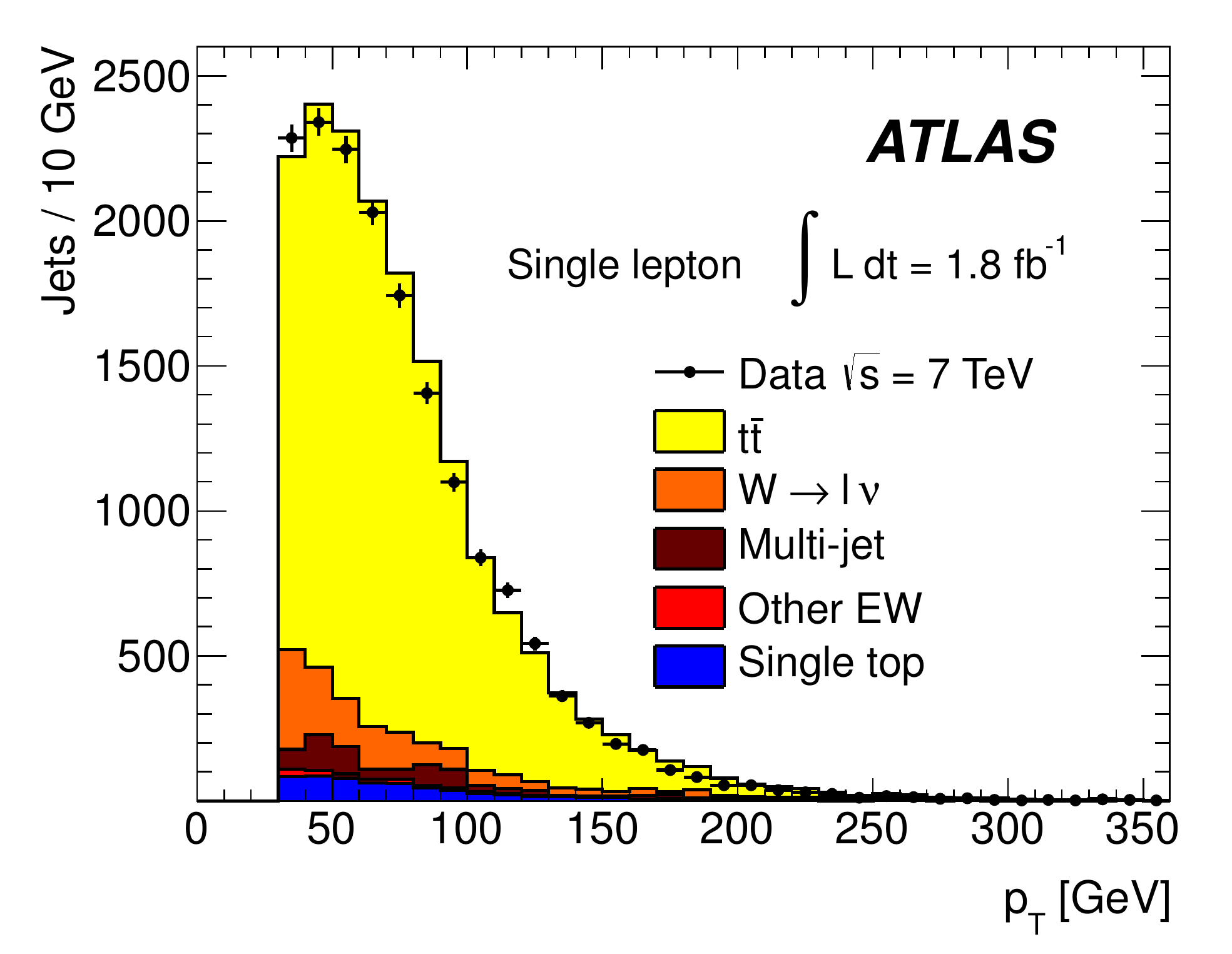}
\hspace{0.7cm}
\includegraphics[width=8.5cm,height=6.0cm]{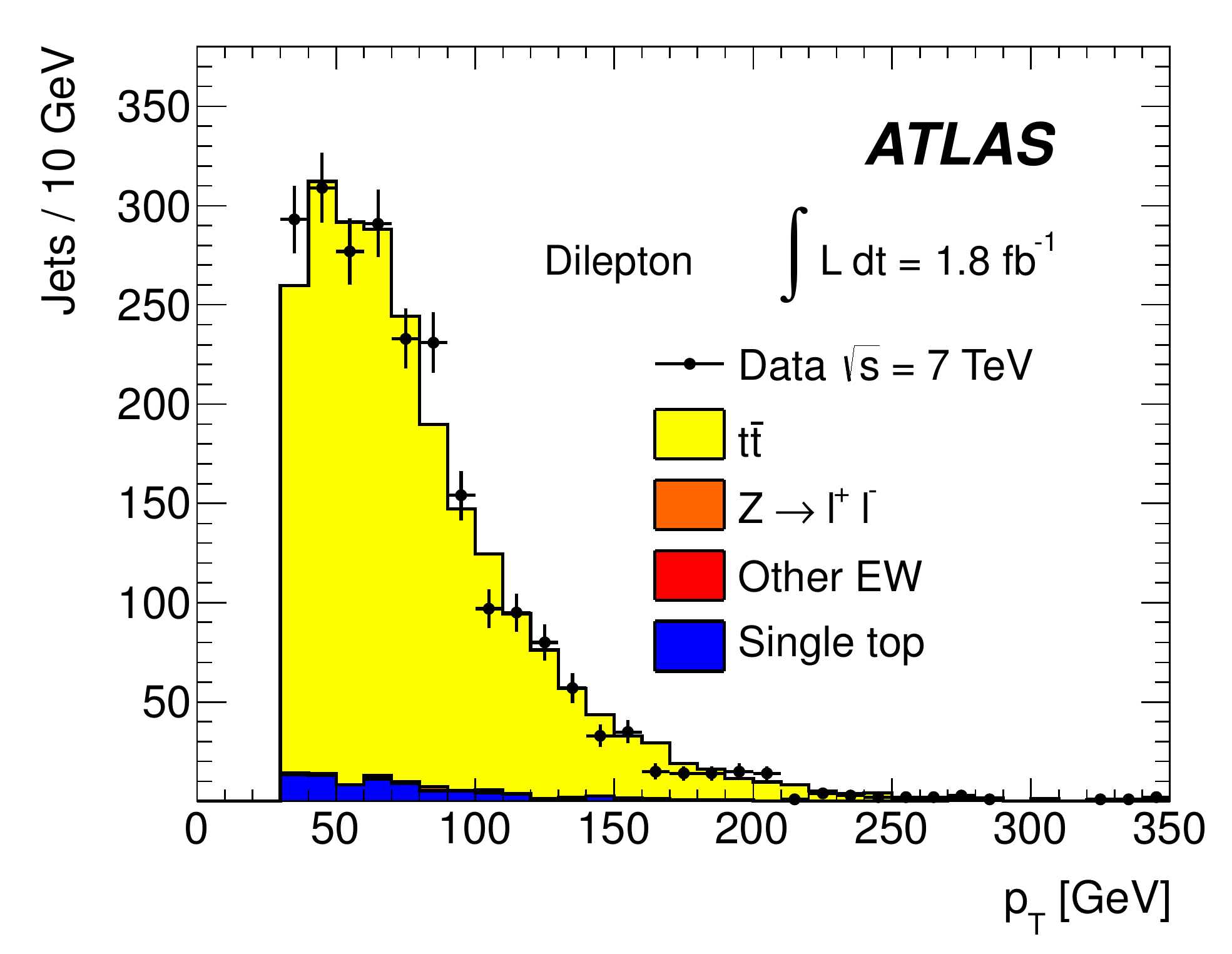}
\caption{The $p_{\mathrm{T}}$ distributions for $b$-tagged jets in the single-lepton (top) and dilepton (bottom) samples along with the sample composition expectations.}
\label{fig:bsample_ds}
\end{figure}
The $p_{\mathrm{T}}$ distributions for the $b$-jets in both the dilepton and single-lepton samples show a similar behaviour, since they come mainly from top-quark decays. This is well described by the MC expectations from the \textsc{MC@NLO} generator coupled to \textsc{Herwig}. In the dilepton sample the signal-to-background ratio is found to be greater than in the single-lepton sample, as it is quantitatively shown in Tables \ref{table1} and \ref{table2}.
\subsection{Light-quark jet sample}
The hadronic decays $\Wboson\rightarrow q\bar{q'}$ are a clean source of light-quark jets, as gluons and $b$-jets are highly suppressed; the former because gluons would originate in radiative corrections of order $\mathcal{O}(\alpha_s)$, and the latter because of the smallness of the CKM matrix elements $\Vub$ and $\Vcb$. To select the light-jet sample, the jet pair in the event which has the invariant mass closest to the $\Wboson$-boson mass is selected. Both jets are also required to be non-tagged by the $b$-tagging algorithm. The number of jets satisfying these criteria is 7158. Figure \ref{fig:lsampleA} shows the transverse momentum distribution of these jets together with the invariant mass of the dijet system.
\begin{figure}
\centering
\includegraphics[width=8.5cm,height=6.0cm]{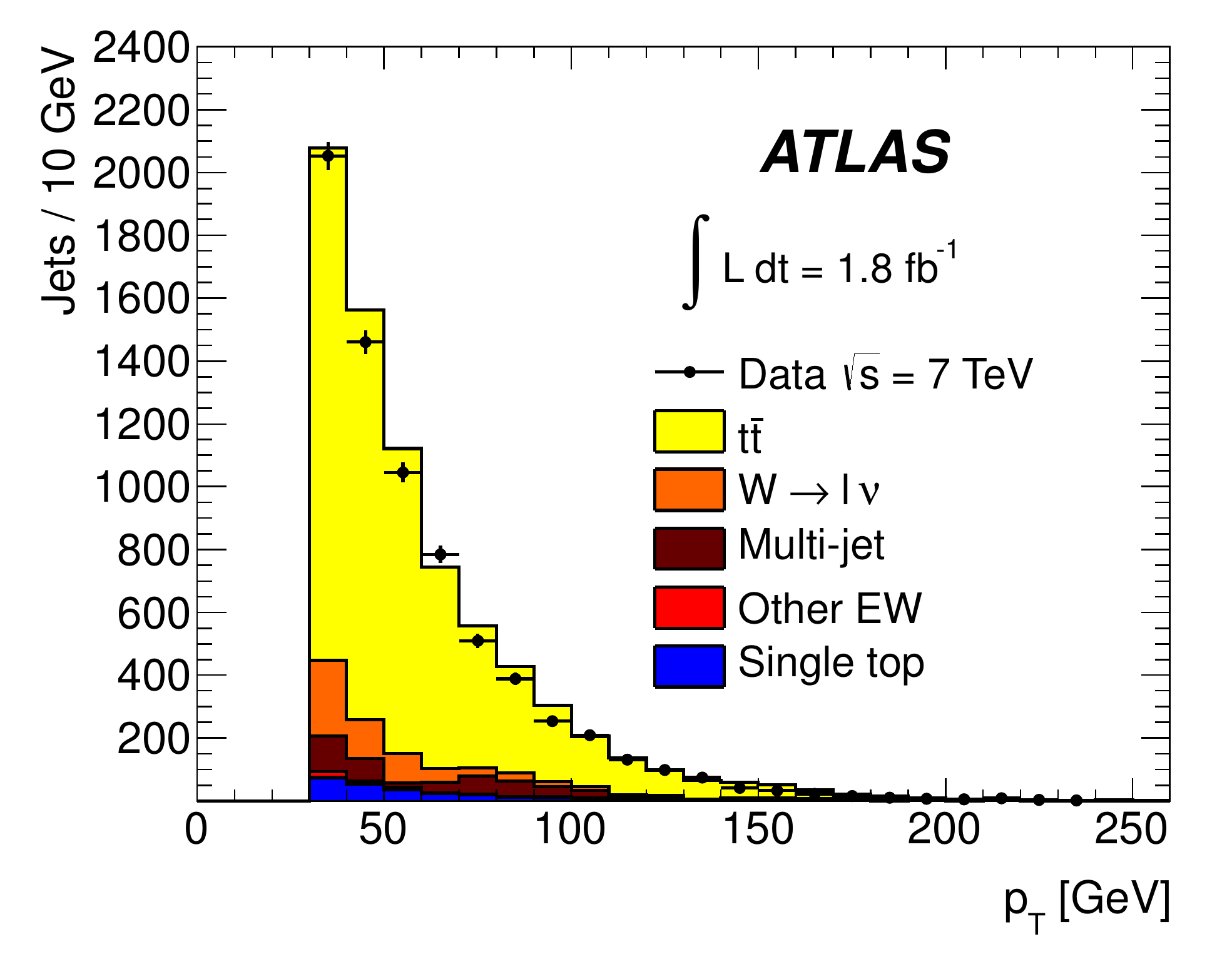}
\hspace{0.7cm}
\includegraphics[width=8.5cm,height=6.0cm]{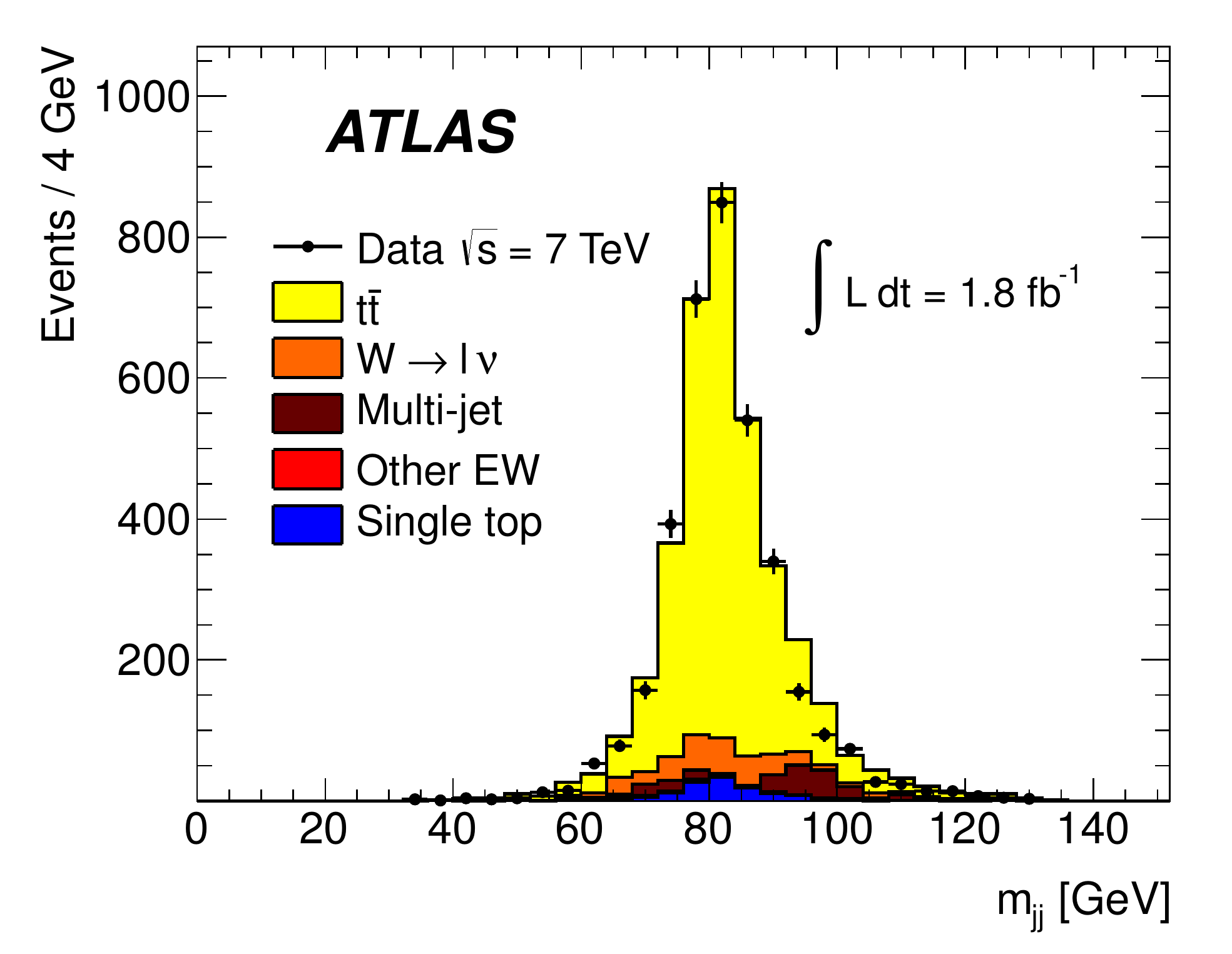}
\caption{The distribution of light-jet $p_{\mathrm{T}}$ (top) and of the invariant mass of light-jet pairs (bottom) along with the sample composition expectations. The latter shows a peak at the $\Wboson$ mass, whose width is determined by the dijet mass resolution.}
\label{fig:lsampleA}
\end{figure}
As expected, the $p_{\mathrm{T}}$ distribution of the light jets from $\Wboson$-boson decays exhibits a stronger fall-off than that for the $b$-jets. 
This dependence is again well described by the MC simulations in the jet $p_{\mathrm{T}}$ region used in this analysis. Agreement between the invariant mass distributions for observed and simulated events is good, in particular in the region close to the $\Wboson$-boson mass.
\subsection{Jet purities}
To estimate the actual number of $b$-jets and light jets in each of the samples, the MC simulation is used by analysing the information at generator level. For $b$-jets, a matching to a $b$-hadron is performed within a radius $\Delta R = 0.3$. For light jets, the jet is required not to have a $b$-hadron within $\Delta R = 0.3$ of the jet axis. Additionally, to distinguish light quarks and $c$-quarks from gluons, the MC parton with highest $p_{\mathrm{T}}$ within the cone of the reconstructed jet is required to be a ($u,d,c$ or $s$)-quark. The purity $p$ is then defined as
\begin{eqnarray}
p = \sum_{k}\alpha_{k}p_{k}; \ \ \ p_{k} = 1-\frac{N_{\mathrm{f}}^{(k)}}{N_{\mathrm{T}}^{(k)}}
\label{eq:pur}
\end{eqnarray}
where $\alpha_k$ is the fraction of events in the $k$-th MC sample (signal or background), given in Tables \ref{table1} and \ref{table2} and $N_{\mathrm{f}}^{(k)}$, $N_{\mathrm{T}}^{(k)}$ are the number of fakes (jets not assigned to the correct flavour, e.g. charm jets in the $b$-jet sample), and the total number of jets in a given sample, respectively. The purity in the multi-jet background is determined using \textsc{Pythia} MC samples.\\
In the single-lepton channel, the resulting purity of the $b$-jet sample is $p^{(\mathrm{s})}_b=(88.5 \pm 5.7) \%$, while the purity of the light-jet sample is found to be $p^{(\mathrm{s})}_{\mathrm{l}}= (66.2 \pm 4.1)\%$, as shown in Table \ref{tablePur}. The uncertainty on the purity arises from the uncertainties on the signal and background fractions in each sample. The charm content in the light-jet sample is found to be 16\%, with the remaining 50\% ascribed to $u,d$ and $s$.\\
MC studies indicate that the contamination of the $b$-jet sample is dominated by charm-jet fakes and that the gluon contamination is about $0.7\%$. For the light-jet sample, the fraction of gluon fakes amounts to $19\%$, while the $b$-jet fakes correspond to $15\%$.\\
In the dilepton channel, a similar calculation yields the purity of the $b$-jet sample to be $p^{(\mathrm{d})}_b = (99.3^{+0.7}_{-6.5})\%$ as shown in Table \ref{tablePur2}. Thus, the $b$-jet sample purity achieved using $t\bar{t}$ final states is much higher than that obtained in inclusive $b$-jet measurements at the Tevatron \cite{CDF} or the LHC \cite{atlasBxsec}.
\begin{table}
\caption{Purity estimation for $b$-jets and light jets in the single-lepton channel. The uncertainty on the purity arises from the uncertainties in the signal and background fractions.}
\label{tablePur}
\begin{center}
\begin{tabular}{cccc}
\hline
Process & $\alpha_k$ & $p_k (b)$ & $p_k$ (light)\\
\hline
$t\bar{t}$ & 0.774 & 0.961 & 0.725\\
$\Wboson\rightarrow \ell\nu$ & 0.128 & 0.430 & 0.360\\
Multi-jet & 0.050 & 0.887 & 0.485\\
Other EW ($\Zboson$, diboson) & 0.011 & 0.611 & 0.342\\
Single top & 0.037 & 0.958 & 0.716\\
\hline
\bf{Weighted total} & - & $(88.5 \pm 5.7)\%$ & $(66.2 \pm 4.1)\%$ \\
\end{tabular}
\end{center}
\end{table}

\begin{table}
\caption{Purity estimation for $b$-jets in the dilepton channel. The uncertainty on the purity arises from the uncertainties in the signal and background fractions.}
\label{tablePur2}
\begin{center}
\begin{tabular}{ccc}
\hline
Process & $\alpha_k$ & $p_k (b)$\\

\hline
$t\bar{t}$ & 0.949 & 0.997\\
$\Zboson \rightarrow \ell^{+}\ell^{-}$& 0.006 & 0.515\\
Other EW ($\Wboson$, diboson) & 0.002 & 0.375\\
Single top & 0.043 & 0.987\\
Multi-jet & - & -\\
\hline
\bf{Weighted total} & - & $(99.3^{+0.7}_{-6.5})\%$\\
\end{tabular}
\end{center}
\end{table}
\section{Jet shapes in the single-lepton channel}
\label{sec7}
For the jet shape calculation, locally calibrated topological clusters are used \cite{JES,lc1,lc2}. In this procedure, effects due to calorimeter response, leakage, and losses in the dead material upstream of the calorimeter are taken into account separately for electromagnetic and hadronic clusters \cite{menke}.\\
The differential jet shape $\rho(r)$ in an annulus of inner radius $r-\Delta r/2$ and outer radius $r+\Delta r/2$ from the axis of a given jet is defined as
\begin{eqnarray}
\rho(r) = \frac{1}{\Delta r}\frac{p_{\mathrm{T}}(r-\Delta r/2,r+\Delta r/2)}{p_{\mathrm{T}}(0,R)}
\end{eqnarray}
Here, $\Delta r = 0.04$ is the width of the annulus; $r$, such that $\Delta r/2 \leq r \leq R-\Delta r/2$, is the distance to the jet axis in the $\eta$-$\phi$ plane, and $p_{\mathrm{T}}(r_1,r_2)$ is the scalar sum of the $p_{\mathrm{T}}$ of the jet constituents with radii between $r_1$ and $r_2$.\\
Some distributions of $\rho (r)$ are shown in Fig. \ref{fig:rhob} for the $b$-jet sample selected in the single-lepton channel.
\begin{figure}
\centering
\includegraphics[width=8.5cm,height=11.0cm]{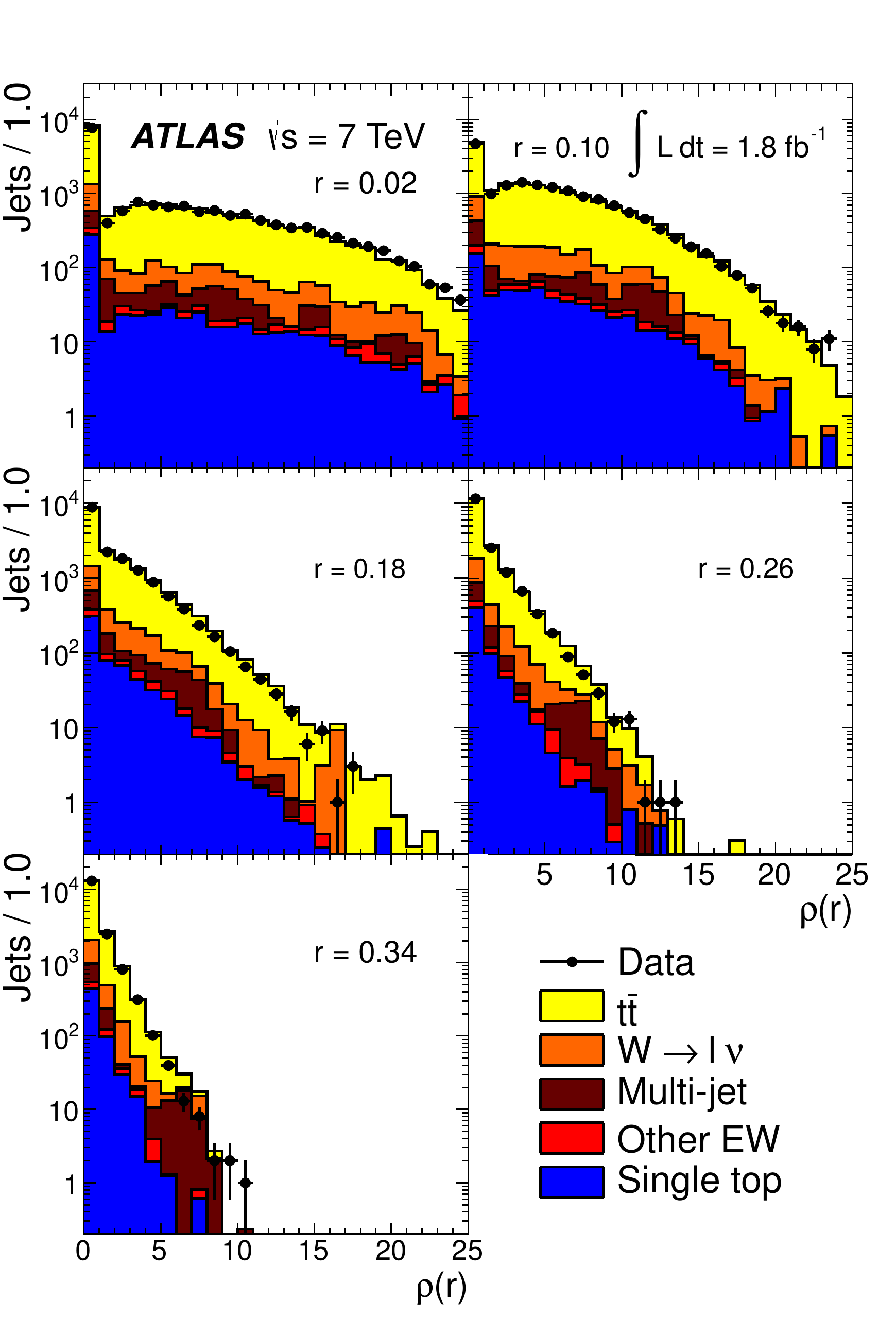}
\caption{Distribution of $R = 0.4$ $b$-jets in the single-lepton channel as a function of the differential jet shapes $\rho(r)$ for different values of $r$.}
\label{fig:rhob}
\end{figure}
There is a marked peak at zero energy deposit, which indicates that energy is concentrated around relatively few particles. As $r$ increases, the distributions of $\rho(r)$ are concentrated at smaller values because of the relatively low energy density at the periphery of the jets. Both effects are well reproduced by the MC generators.\\
The analogous $\rho(r)$ distributions for light jets are shown in Fig. \ref{fig:rhol}. The gross features are similar to those previously discussed for $b$-jets, but for small values of $r$, the $\rho(r)$ distributions for light jets are somewhat flatter than those for $b$-jets.\\
\begin{figure}
\centering
\includegraphics[width=8.5cm,height=11.0cm]{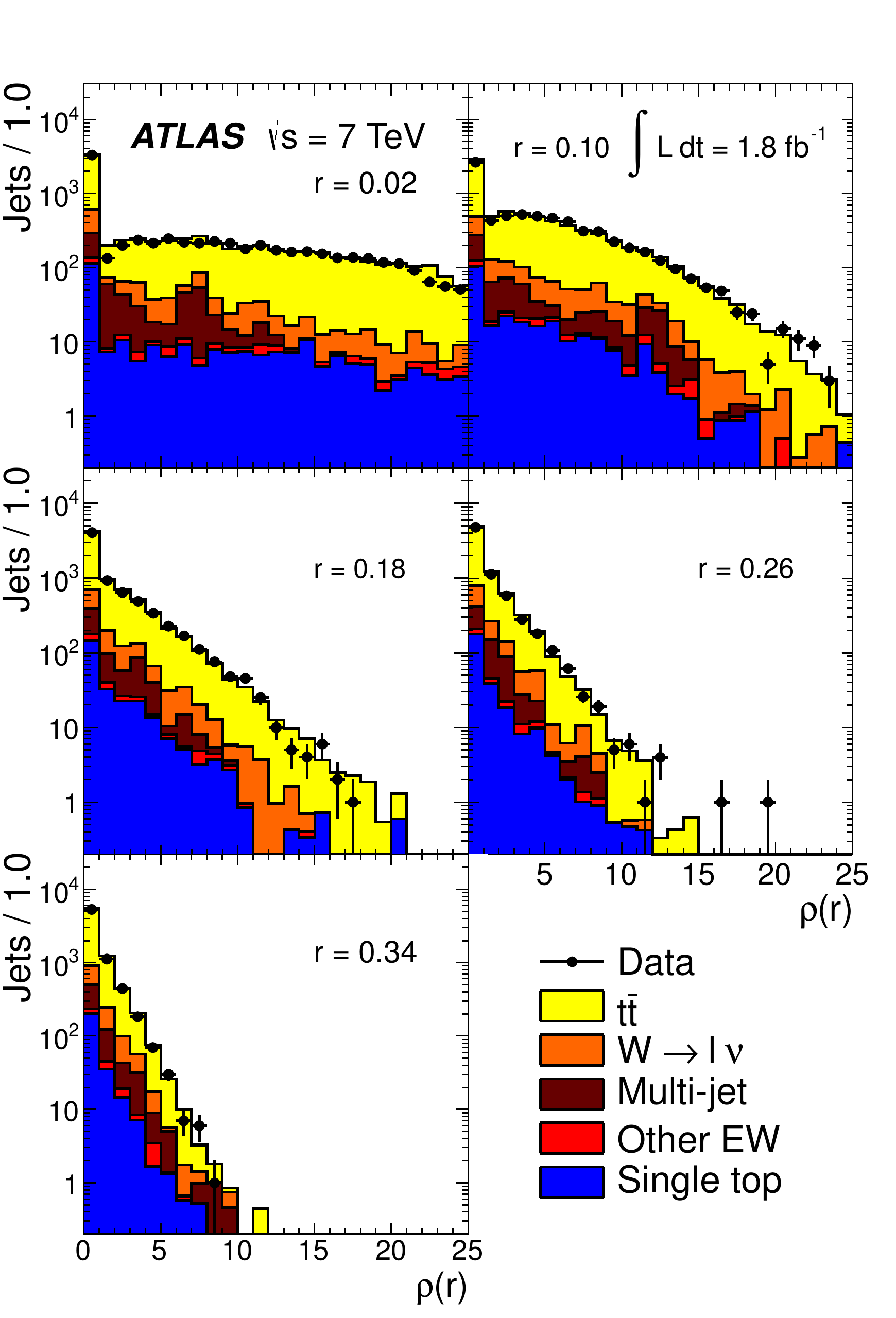}
\caption{Distribution of $R = 0.4$ light jets in the single-lepton channel as a function of the differential jet shapes $\rho(r)$ for different values of $r$.}
\label{fig:rhol}
\end{figure}
The integrated jet shape in a cone of radius $r \leq R$ around the jet axis is defined as the cumulative distribution for $\rho(r)$, i.e.
\begin{eqnarray}
\Psi(r) = \frac{p_{\mathrm{T}}(0,r)}{p_{\mathrm{T}}(0,R)}; \ \ r \leq R
\end{eqnarray}
which satisfies $\Psi(r = R) = 1$. Figure \ref{fig:psib} (\ref{fig:psil}) shows distributions of the integrated jet shapes for $b$-jets (light jets) in the single-lepton sample.
\begin{figure}
\centering
\includegraphics[width=8.5cm,height=11.0cm]{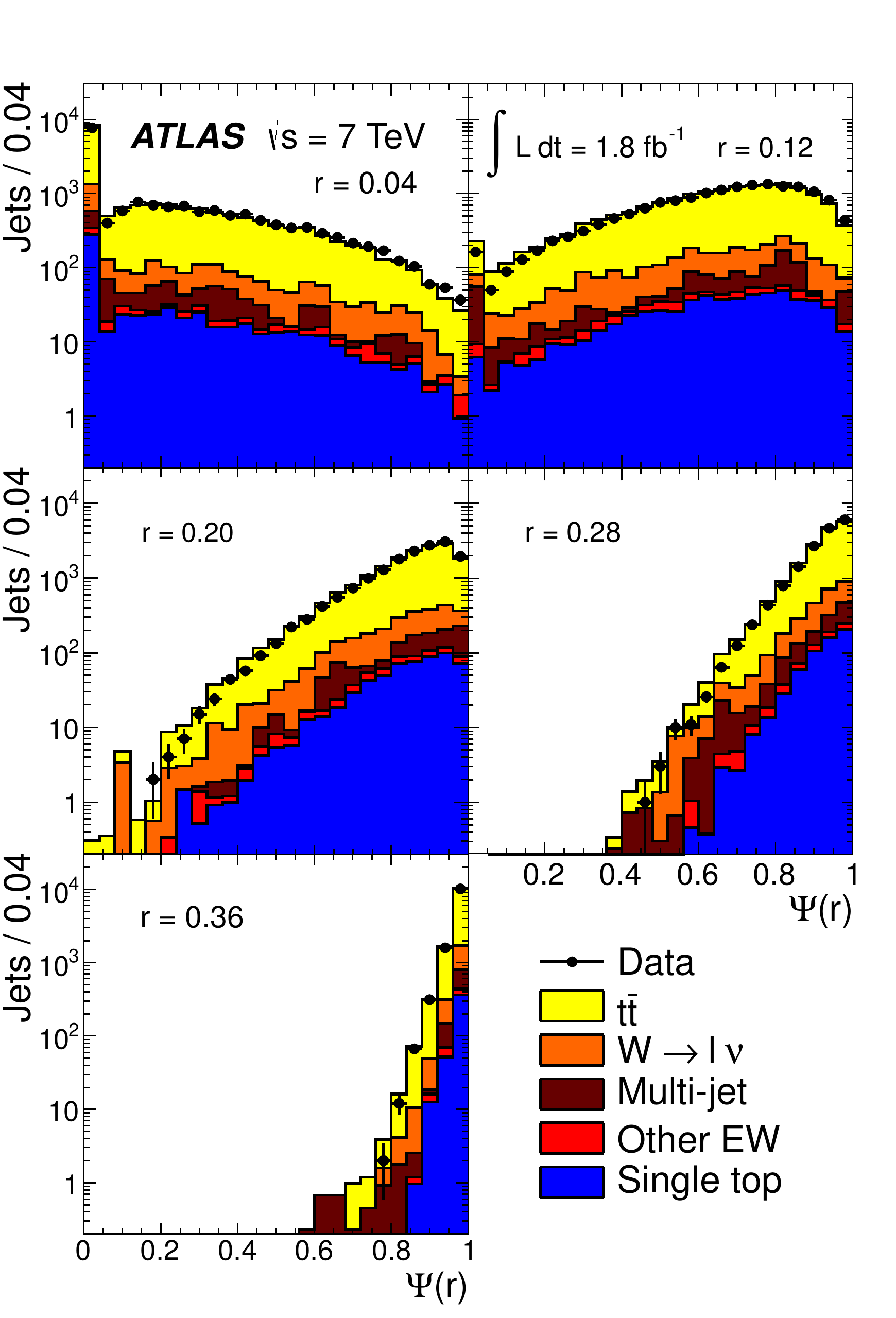}
\caption{Distribution of $R = 0.4$ $b$-jets in the single-lepton channel as a function of the integrated jet shapes $\Psi(r)$ for different values of $r$.}
\label{fig:psib}
\end{figure}
\begin{figure}
\centering
\includegraphics[width=8.5cm,height=11.0cm]{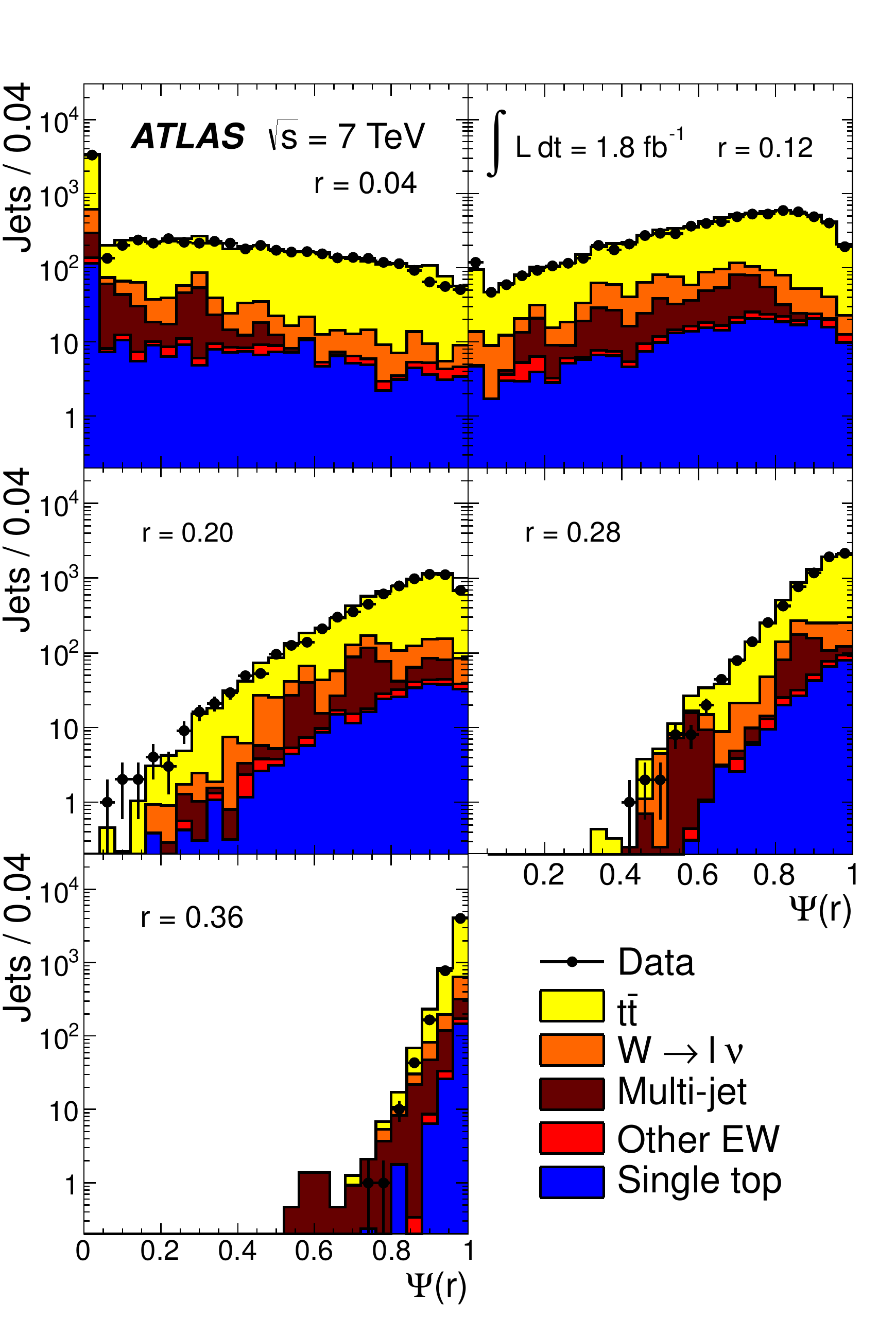}
\caption{Distribution of $R = 0.4$ light jets in the single-lepton channel as a function of the integrated jet shapes $\Psi(r)$ for different values of $r$.}
\label{fig:psil}
\end{figure}
These figures show the inclusive (i.e. not binned in either $\eta$ or $p_{\mathrm{T}}$) $\rho (r)$ and $\Psi(r)$ distributions for fixed values of $r$. Jet shapes are only mildly dependent on pseudorapidity, while they strongly depend on the transverse momentum. This behaviour has been verified in previous analyses \cite{d0,h1,hera2,chek,cdf01,lhc,cms}. This is illustrated in Figs. \ref{fig:shapePt} and \ref{fig:shapeEta}, which show the energy fraction in the outer half of the cone as a function of $p_{\mathrm{T}}$ and $|\eta|$. For this reason, all the data presented in the following are binned in five $p_{\mathrm{T}}$ regions with $p_{\mathrm{T}} < 150 \GeV$, where the statistical uncertainty is small enough.
In the following, only the average values of these distributions are presented:
\begin{figure}
\centering
\includegraphics[width=8.5cm,height=6.0cm]{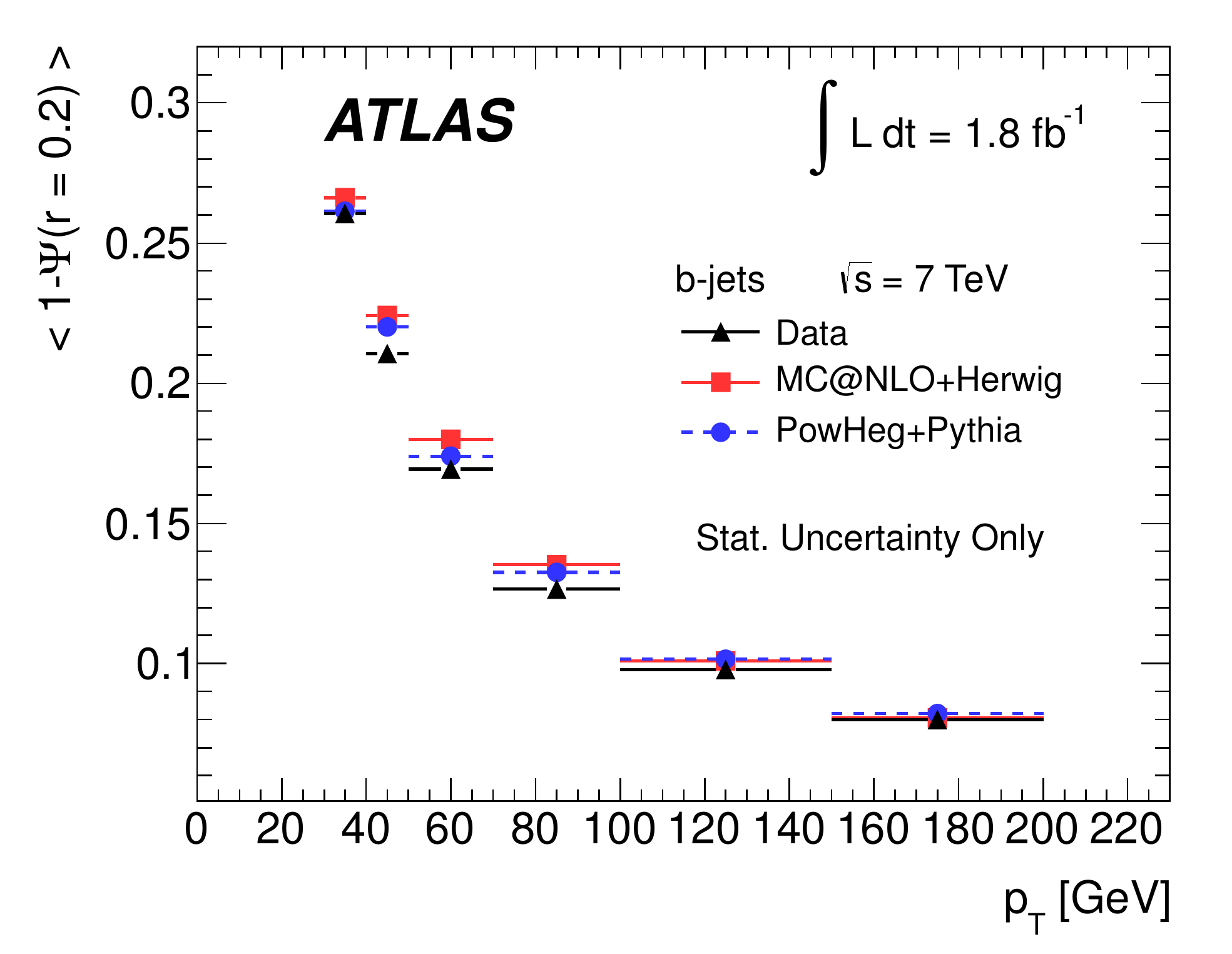}
\includegraphics[width=8.5cm,height=6.0cm]{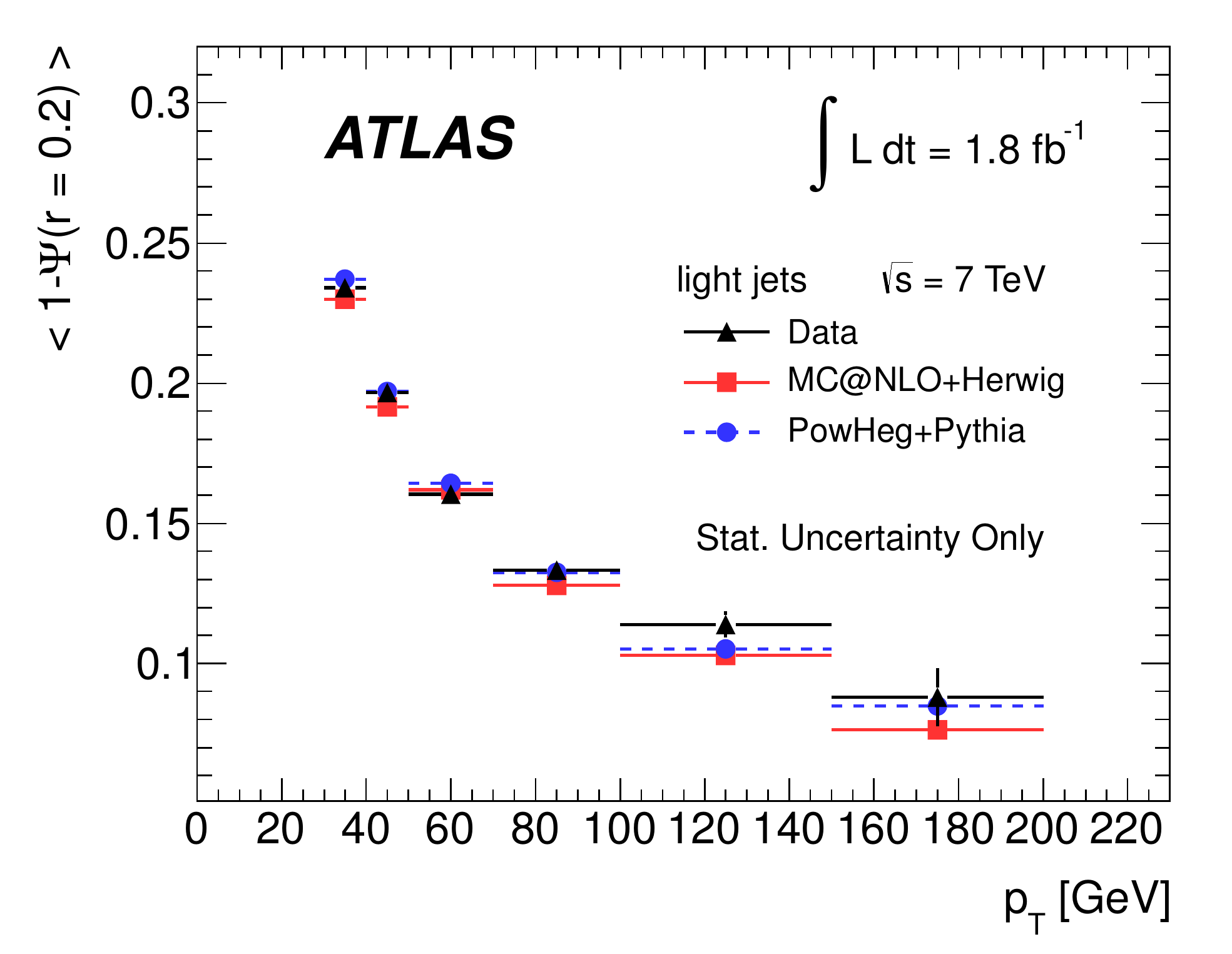}
\caption{Dependence of the $b$-jet (top) and light-jet (bottom) shapes on the jet transverse momentum. This dependence is quantified by plotting the mean value $\langle 1-\Psi(r = 0.2)\rangle$ (the fraction of energy in the outer half of the jet cone) as a function of $p_{\mathrm{T}}$ for jets in the single-lepton sample.}
\label{fig:shapePt}
\end{figure}

\begin{figure}
\centering
\includegraphics[width=8.5cm,height=6.0cm]{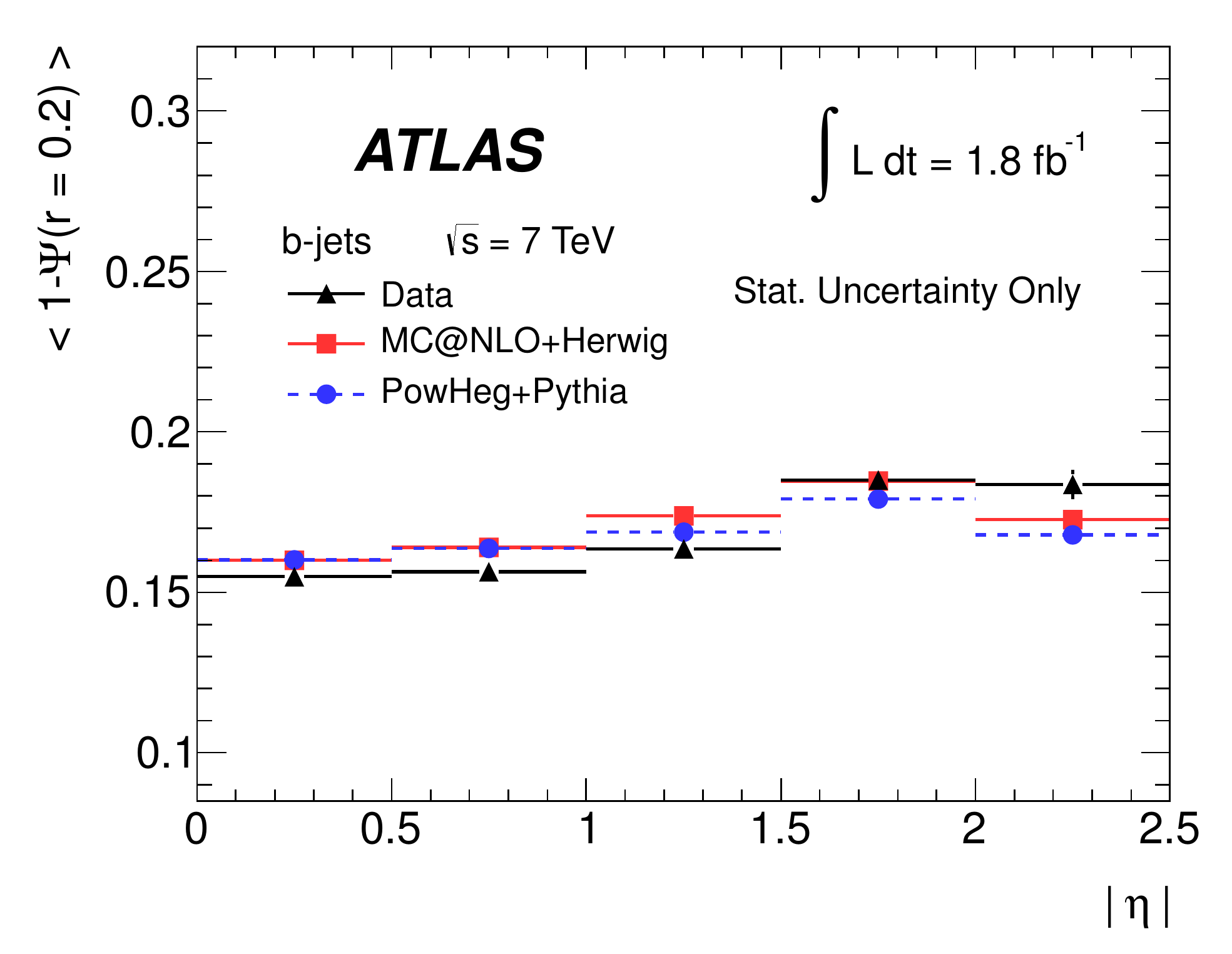}
\includegraphics[width=8.5cm,height=6.0cm]{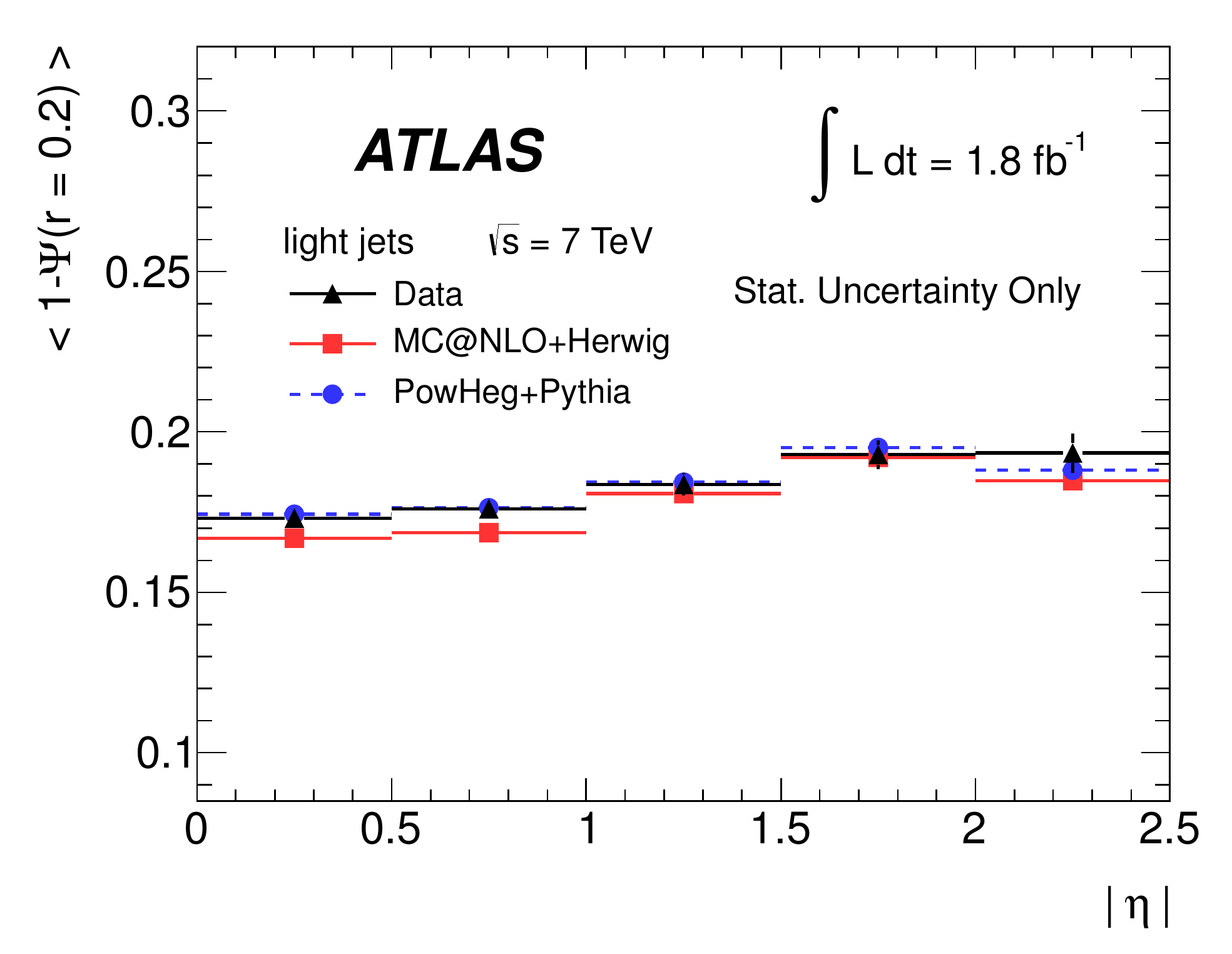}
\caption{Dependence of the $b$-jet (top) and light-jet (bottom) shape on the jet pseudorapidity. This dependence is quantified by plotting the mean value $\langle 1-\Psi(r = 0.2)\rangle$ (the fraction of energy in the outer half of the jet cone) as a function of $|\eta|$ for jets in the single-lepton sample.}
\label{fig:shapeEta}
\end{figure}
\begin{eqnarray}
\langle\rho(r)\rangle & = & \frac{1}{\Delta r}\frac{1}{N_{\mathrm{jets}}}\sum_{\mathrm{jets}}\frac{p_{\mathrm{T}}(r-\Delta r/2,r+\Delta r/2)}{p_{\mathrm{T}}(0,R)}\\
\langle\Psi(r)\rangle & = & \frac{1}{N_{\mathrm{jets}}}\sum_{\mathrm{jets}}\frac{p_{\mathrm{T}}(0,r)}{p_{\mathrm{T}}(0,R)}
\label{meanShape}
\end{eqnarray}
where the sum is performed over all jets of a given sample, light jets ($l$) or $b$-jets ($b$) and $N_{\mathrm{jets}}$ is the number of jets in the sample.
\section{Results at the detector level}
\label{sec8}
In the following, the detector-level results for the average values $\langle \rho(r) \rangle$ and $\langle \Psi(r) \rangle$ as a function of the jet internal radius $r$, are presented. A comparison has been made between $b$-jet shapes obtained in both the dilepton and single-lepton samples, and it is found that they are consistent with each other within the uncertainties. Thus the samples are merged. In Figure \ref{fig:detRho}, the distributions for the average values of the differential jet shapes are shown for each $p_{\mathrm{T}}$ bin, along with a comparison with the expectations from the simulated samples described in Sect. \ref{sec3}.
\begin{figure}
\centering
\includegraphics[width=8.5cm,height=12.5cm]{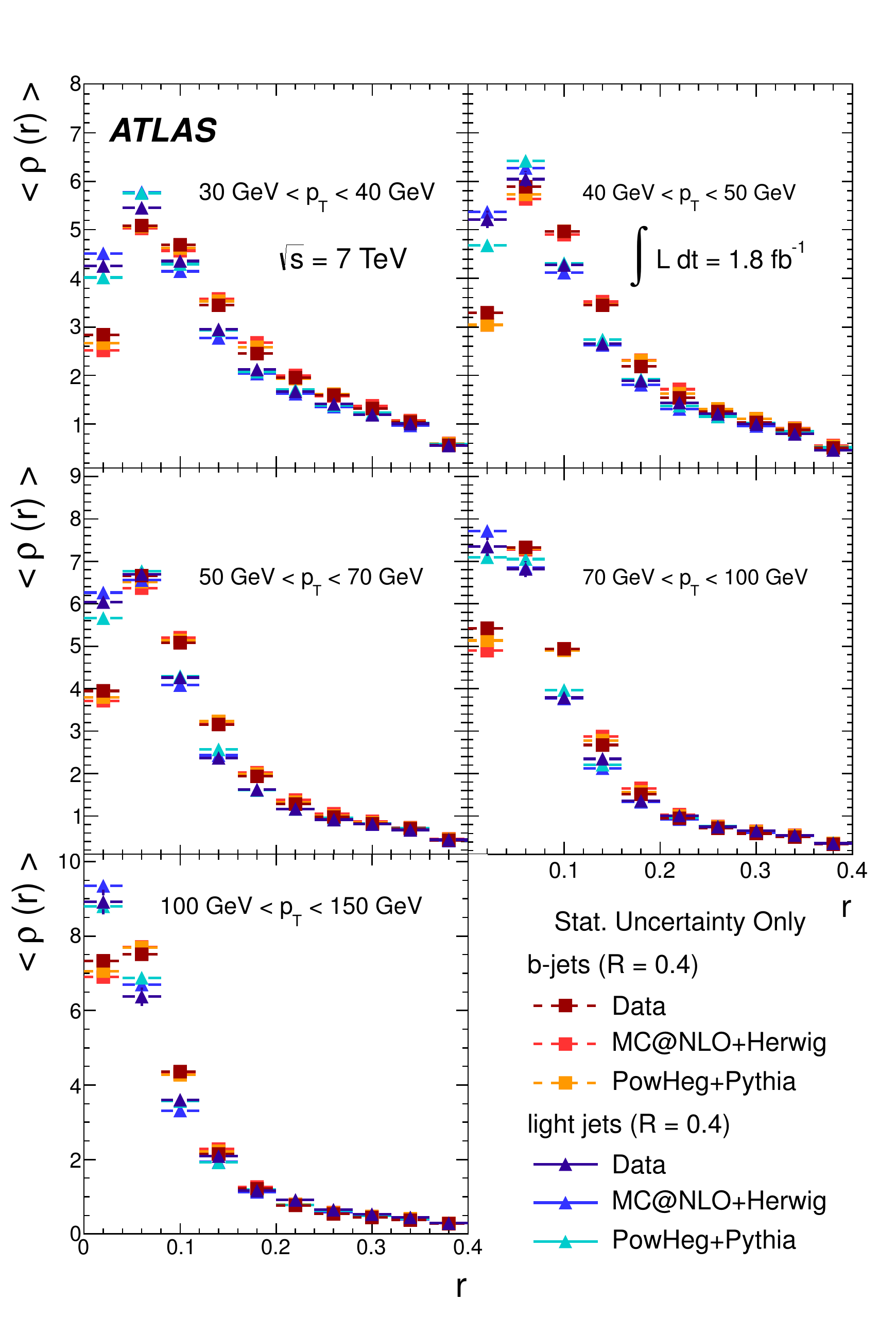}
\caption{Average values of the differential jet shapes $\langle\rho(r)\rangle$ for light jets (triangles) and $b$-jets (squares), with $\Delta r = 0.04$, as a function of $r$ at the detector level, compared to \textsc{MC@NLO+Herwig} and \textsc{Powheg+Pythia} event generators. The uncertainties shown for data are only statistical.}
\label{fig:detRho}
\end{figure}
There is a small but clear difference between light- and $b$-jet differential shapes, the former lying above (below) the latter for smaller (larger) values of $r$. These differences are more visible at low transverse momentum. In Figure \ref{fig:detPsi}, the average integrated jet shapes $\langle\Psi(r)\rangle$ are shown for both the light jets and $b$-jets, and compared to the MC expectations discussed earlier. Similar comments apply here:
\begin{figure}
\centering
\includegraphics[width=8.5cm,height=12.5cm]{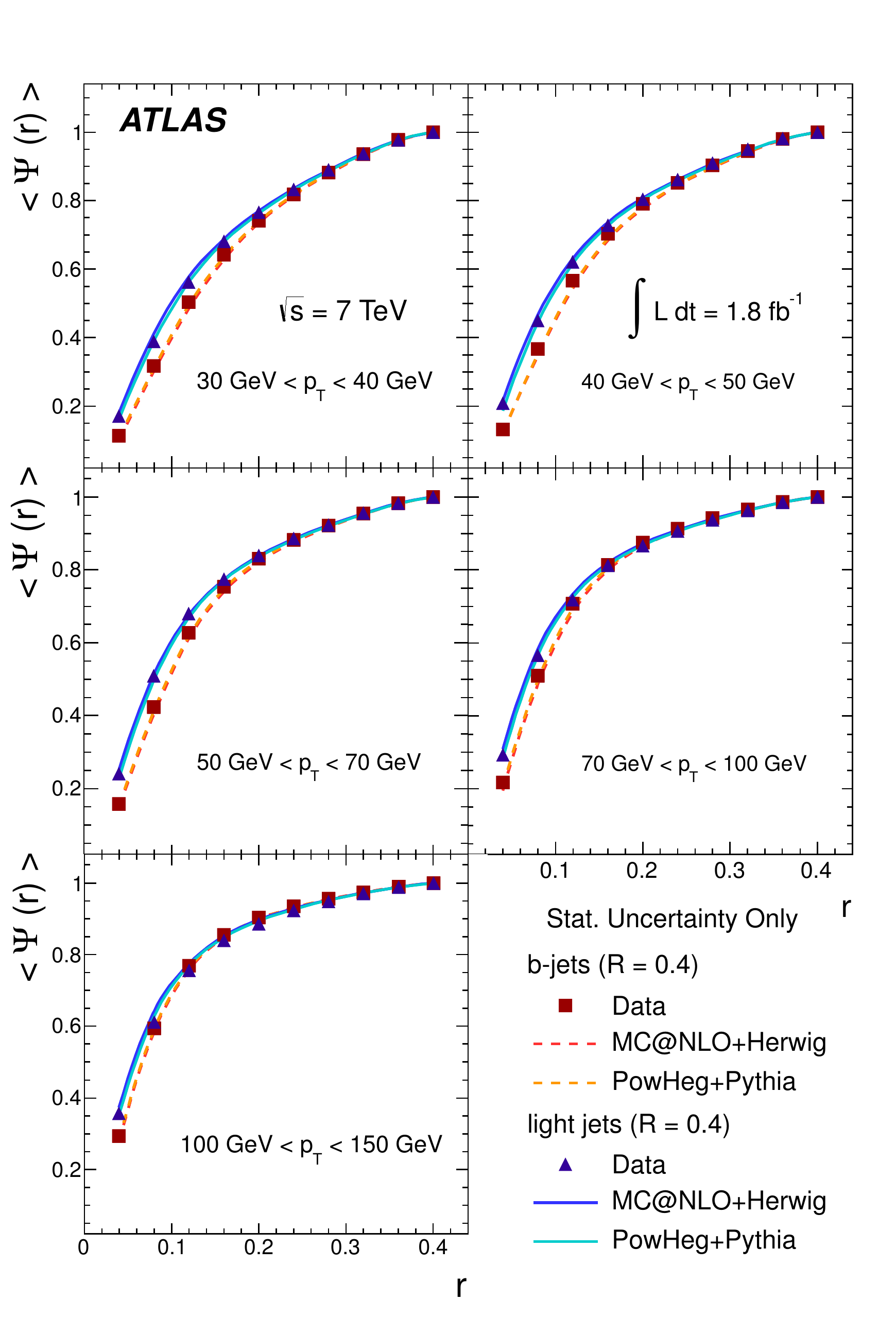}
\caption{Average values of the integrated jet shapes $\langle\Psi(r)\rangle$ for light jets (triangles) and $b$-jets (squares), with $\Delta r = 0.04$, as a function of $r$ at the detector level, compared to \textsc{MC@NLO+Herwig} and \textsc{Powheg+Pythia} event generators. The uncertainties shown for data are only statistical.}
\label{fig:detPsi}
\end{figure}
The values of $\langle\Psi(r)\rangle$ are consistently larger for light jets than for $b$-jets for small values of $r$, while they tend to merge as $r \rightarrow R$ since, by definition, $\Psi(R) = 1$.
\section{Unfolding to particle level}
\label{sec9}
In order to correct the data for acceptance and detector effects, thus enabling comparisons with different models and other experiments, an unfolding procedure is followed. The method used to correct the measurements based on topological clusters to the particle level relies on a bin-by-bin correction. Correction factors $F(r)$ are calculated separately for differential, $\langle\rho(r)\rangle$, and integrated, $\langle\Psi(r)\rangle$, jet shapes in both the light- and $b$-jet samples. For differential ($\rho$) and integrated jet shapes ($\Psi$), they are defined as the ratio of the particle-level quantity to the detector-level quantity as described by the MC simulations discussed in Sect. \ref{sec3}, i.e.
\begin{eqnarray}
F^{\rho}_{\mathrm{l},b}(r) = \frac{\langle\rho(r)_{\mathrm{l},b}\rangle_{\mathrm{MC,part}}} {\langle\rho(r)_{\mathrm{l},b}\rangle_{\mathrm{MC,det}}}\\
F^{\Psi}_{\mathrm{l},b}(r) = \frac{\langle\Psi(r)_{\mathrm{l},b}\rangle_{\mathrm{MC,part}}} {\langle\Psi(r)_{\mathrm{l},b}\rangle_{\mathrm{MC,det}}}
\end{eqnarray}
While the detector-level MC includes the background sources described before, the particle-level jets are built using all particles in the signal sample with an average lifetime above $10^{-11} \mathrm{s}$, excluding muons and neutrinos. The results have only a small sensitivity to the inclusion or not of muons and neutrinos, as well as to the background estimation. For particle-level $b$-jets, a $b$-hadron with $p_{\mathrm{T}} > 5 \GeV$ is required to be closer than $\Delta R = 0.3$ from the jet axis, while for light jets, a selection equivalent to that for the detector-level jets is applied, selecting the non-$b$-jet pair with invariant mass closest to $m_{\Wboson}$. The same kinematic selection criteria are applied to these particle-level jets as for the reconstructed jets, namely $p_{\mathrm{T}} > 25 \GeV$, $|\eta| < 2.5$ and $\Delta R > 0.8$ to avoid jet--jet overlaps.\\
A Bayesian iterative unfolding approach \cite{bayes} is used as a cross-check. The RooUnfold software \cite{roounfold} is used by providing the jet-by-jet information on the jet shapes, in the $p_{\mathrm{T}}$ intervals defined above. This method takes into account bin-by-bin migrations in the $\rho(r)$ and $\Psi(r)$ distributions for fixed values of $r$. The results of the bin-by-bin and the Bayesian unfolding procedures agree at the 2\% level.\\
As an additional check of the stability of the unfolding procedure, the directly unfolded integrated jet shapes are compared with those obtained from integrating the unfolded differential distributions. The results agree to better than 1\%.
These results are reassuring since the differential and integrated jet shapes are subject to migration and resolution effects in different ways. Both quantities are also subject to bin-to-bin correlations. For the differential measurement, the correlations arise from the common normalisation. They increase with the jet transverse momentum, varying from 25\% to 50\% at their maximum, which is reached for neighbouring bins at low $r$. The correlations for the integrated measurement are greater and their maximum varies from 60\% to 75\% as the jet $\pt$ increases.
\section{Systematic uncertainties}
\label{sec10}
The main sources of systematic uncertainty are described below.
\begin{itemize}
\item{} The energy of individual clusters inside the jet is varied according to studies using isolated tracks \cite{singleHadron}, parameterising the uncertainty on the calorimeter energy measurements as a function of the cluster $p_{\mathrm{T}}$. The impact on the differential jet shape increases from 2\% to 10\% as the edge of the jet cone is approached.
\item{} The coordinates $\eta$, $\phi$ of the clusters are smeared using a Gaussian distribution with an RMS width of 5~mrad accounting for small differences in the cluster position between data and Monte Carlo \cite{boostedJets}. This smearing has an effect on the jet shape which is smaller than 2\%. 
\item{} An uncertainty arising from the amount of passive material in the detector is derived using the algorithm described in Ref. \cite{boostedJets} as a result of the studies carried out in Ref. \cite{singleHadron}. Low-energy clusters ($E < 2.5 \GeV$) are removed from the reconstruction according to a probability function given by $\mathcal{P}(E=0)\times \mathrm{e}^{-2E}$, where $\mathcal{P}(E = 0)$ is the measured probability (28\%) of a charged particle track to be associated with a zero energy deposit in the calorimeter and $E$ is the cluster energy in $\GeV$. As a result, approximately 6\% of the total number of clusters are discarded. The impact of this cluster-removing algorithm on the measured jet shapes is smaller than 2\%.
\item{} As a further cross-check an unfolding of the track-based jet shapes to the particle level has also been performed. The differences from those obtained using calorimetric measurements are of a similar scale to the ones discussed for the cluster energy, angular smearing and dead material.
\item{} An uncertainty arising from the jet energy calibration (JES) is taken into account by varying the jet energy scale in the range 2\% to 8\% of the measured value, depending on the jet $p_{\mathrm{T}}$ and $\eta$. This variation is different for light jets and $b$-jets since they have a different particle composition. 
\item{} The jet energy resolution is also taken into account by smearing the jet $p_{\mathrm{T}}$ using a Gaussian centred at unity and with standard deviation $\sigma_{\mathrm{r}}$ \cite{jetResol}. The impact on the measured jet shapes is about 5\%.
\item{} The uncertainty due to the JVF requirement is estimated by comparing the jet shapes with and without this requirement. The uncertainty is smaller than 1\%.
\item{} An uncertainty is also assigned to take pile-up effects into account. This is done by calculating the differences between samples where the number of $pp$ interaction vertices is smaller (larger) than five and the total sample. The impact on the differential jet shapes varies from 2\% to 10\% as $r$ increases.
\item{} An additional uncertainty due to the unfolding method is determined by comparing the correction factors obtained with three different MC samples, \textsc{Powheg + Pythia}, \textsc{Powheg + Jimmy} and \textsc{AcerMC} \cite{acer} with the \textsc{Perugia 2010} tune \cite{Perugia}, to the nominal correction factors from the \textsc{MC@NLO} sample. The uncertainty is defined as the maximum deviation of these three unfolding results, and it varies from 1\% to 8\%.
\end{itemize}
Additional systematic uncertainties associated with details of the analysis such as the working point of the $b$-tagging algorithm and the $\Delta R > 0.8$ cut between jets, as well as those related to physics object reconstruction efficiencies and variations in the background normalisation are found to be negligible.
All sources of systematic uncertainty are propagated through the unfolding procedure. The resulting systematic uncertainties on each differential or integrated shape are added in quadrature. 
In the case of differential jet shapes, the uncertainty varies from 1\% to 20\% in each $p_{\mathrm{T}}$ bin as $r$ increases, while the uncertainty for the integrated shapes decreases from 10\% to 0\% as one approaches the edge of the jet cone, where $r = R$.
\section{Discussion of the results}
\label{sec11}
The results at the particle level are presented, together with the total uncertainties arising from statistical and systematic effects. The averaged differential jet shapes $\langle\rho(r)\rangle$ are shown in the even-numbered Figs. \ref{fig:diff1}--\ref{fig:diff5} as a function of $r$ and in bins of $\pt$, while numerical results are presented in the odd-numbered Tables \ref{tabDiff1}--\ref{tabDiff5}. The observation made at the detector level in Sect. \ref{sec8} that $b$-jets are broader than light jets is strengthened after unfolding because it also corrects the light-jet sample for purity effects.
Similarly, the odd-numbered Figs. \ref{fig:int1}--\ref{fig:int5} show the integrated shapes $\langle\Psi(r)\rangle$ as a function of $r$ and in bins of $\pt$ for light jets and $b$-jets. Numerical results are presented in the even-numbered Tables \ref{tabInt1}--\ref{tabInt5}. As before, the observation is made that $b$-jets have a wider energy distribution inside the jet cone than light jets, as it can be seen that $\langle\Psi_b\rangle < \langle\Psi_\mathrm{l}\rangle$ for low $p_{\mathrm{T}}$ and small $r$.\\
These observations are in agreement with the MC calculations, where top-quark pair-production cross sections are implemented using matrix elements calculated to NLO accuracy, which are then supplemented by angular- or transverse momentum-ordered parton showers. Within this context, both \textsc{MC@NLO} and \textsc{Powheg+Pythia} give a good description of the data, as illustrated in Fig. \ref{fig:diff1}--\ref{fig:int5}.\\
Comparisons with other MC approaches have been made (see Fig. \ref{fig:mcComps}).
The \textsc{Perugia 2011} tune, coupled to \textsc{Alpgen+Pythia}, \textsc{Powheg+Pythia} and \textsc{AcerMC+Pythia}, has been compared to the data, and found to be slightly disfavoured. The \textsc{AcerMC} generator \cite{acer} coupled to \textsc{Pythia} for the parton shower and with the \textsc{Perugia 2010} tune \cite{Perugia} gives a somewhat better description of the data, as does the \textsc{Alpgen} \cite{alpgen} generator coupled to \textsc{Herwig}.\\
\textsc{AcerMC} coupled to \textsc{Tune A Pro} \cite{tuneA1,tuneA2} is found to give the best description of the data within the tunes investigated. Colour reconnection effects, as implemented in \textsc{Tune A CR Pro} \cite{tuneA1, tuneA2} have a small impact on this observable, compared to the systematic uncertainties.\\
Since jet shapes are dependent on the method chosen to match parton showers to the matrix-element calculations and, to a lesser extent, on the fragmentation and underlying-event modelling, the measurements presented here provide valuable inputs to constrain present and future MC models of colour radiation in $t\bar{t}$ final states.\\
MC generators predict jet shapes to depend on the hard scattering processs. MC studies were carried out and it was found that inclusive $b$-jet shapes, obtained from the underlying hard processes $gg\to b\bar{b}$ and $gb \to gb$ with gluon splitting $g\to b\bar{b}$ included in the subsequent parton shower, are wider than those obtained in the $t\bar{t}$ final states.
The differences are interpreted as due to the different colour flows in the two different final states i.e. $t\bar{t}$ and inclusive multi-jet production. Similar differences are also found for light-jet shapes, with jets generated in inclusive multi-jet samples being wider than those from $\Wboson$-boson decays in top-quark pair-production. 
\section{Summary}
\label{sec12}
The structure of jets in $t\bar{t}$ final states has been studied in both the dilepton and single-lepton modes using the ATLAS detector at the LHC. The first sample proves to be a very clean and copious source of $b$-jets, as the top-quark decays predominantly via $t\to \Wboson b$. The second is also a clean source of light jets produced in the hadronic decays of one of the $\Wboson$ bosons in the final state. The differences between the $b$-quark and light-quark jets obtained in this environment have been studied in terms of the differential jet shapes $\rho(r)$ and integrated jet shapes $\Psi(r)$. These variables exhibit a marked (mild) dependence on the jet transverse momentum (pseudorapidity).\\
The results show that the mean value $\langle\Psi(r)\rangle$ is smaller for $b$-jets than for light jets in the region where it is possible to distinguish them, i.e. for low values of the jet internal radius $r$. This means that $b$-jets are broader than light-quark jets, and therefore the cores of light jets have a larger energy density than those of $b$-jets. The jet shapes are well reproduced by current MC generators for both light and $b$-jets.

\newpage

\small{
\paragraph{\bf{\small{Acknowledgements}}}{We thank CERN for the very successful operation of the LHC, as well as the
support staff from our institutions without whom ATLAS could not be
operated efficiently.\\
We acknowledge the support of ANPCyT, Argentina; YerPhI, Armenia; ARC,
Australia; BMWF and FWF, Austria; ANAS, Azerbaijan; SSTC, Belarus; CNPq and FAPESP,
Brazil; NSERC, NRC and CFI, Canada; CERN; CONICYT, Chile; CAS, MOST and NSFC,
China; COLCIENCIAS, Colombia; MSMT CR, MPO CR and VSC CR, Czech Republic;
DNRF, DNSRC and Lundbeck Foundation, Denmark; EPLANET, ERC and NSRF, European Union;
IN2P3-CNRS, CEA-DSM/ IRFU, France; GNSF, Georgia; BMBF, DFG, HGF, MPG and AvH
Foundation, Germany; GSRT and NSRF, Greece; ISF, MINERVA, GIF, DIP and Benoziyo Center,
Israel; INFN, Italy; MEXT and JSPS, Japan; CNRST, Morocco; FOM and NWO,
Netherlands; BRF and RCN, Norway; MNiSW, Poland; GRICES and FCT, Portugal; MERYS
(MECTS), Romania; MES of Russia and ROSATOM, Russian Federation; JINR; MSTD,
Serbia; MSSR, Slovakia; ARRS and MIZ\v{S}, Slovenia; DST/NRF, South Africa;
MICINN, Spain; SRC and Wallenberg Foundation, Sweden; SER, SNSF and Cantons of
Bern and Geneva, Switzerland; NSC, Taiwan; TAEK, Turkey; STFC, the Royal
Society and Leverhulme Trust, United Kingdom; DOE and NSF, United States of
America.\\
The crucial computing support from all WLCG partners is acknowledged
gratefully, in particular from CERN and the ATLAS Tier-1 facilities at
TRIUMF (Canada), NDGF (Denmark, Norway, Sweden), CC-IN2P3 (France),
KIT/GridKA (Germany), INFN-CNAF (Italy), NL-T1 (Netherlands), PIC (Spain),
ASGC (Taiwan), RAL (UK) and BNL (USA) and in the Tier-2 facilities
worldwide.}}

\clearpage

\begin{figure}[]
\vspace{0.8cm}
\includegraphics[width=8.5cm,height=10.5cm]{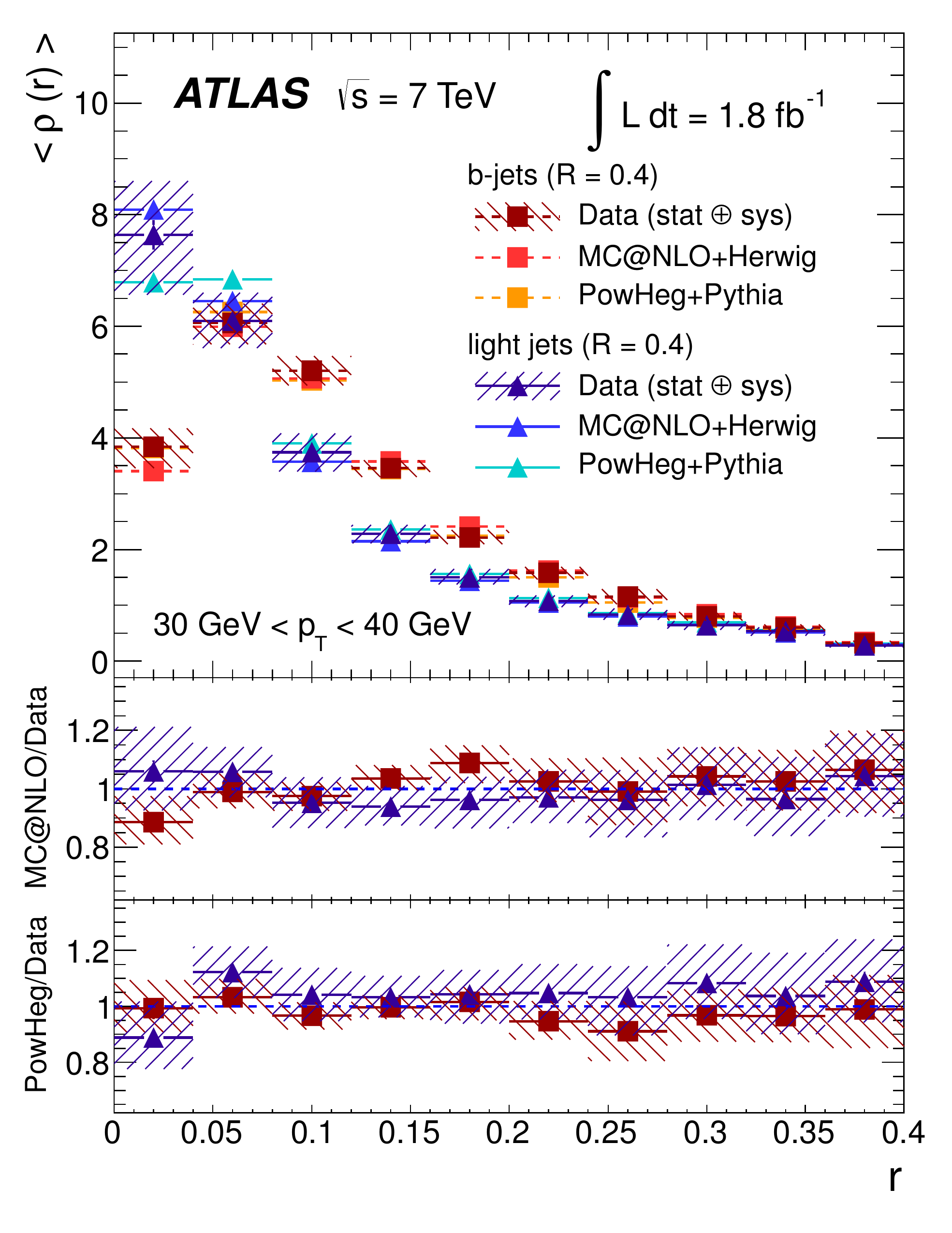}
\caption{Differential jet shapes $\langle\rho(r)\rangle$ as a function of the radius $r$ for light jets (triangles) and $b$-jets (squares). The data are compared to \textsc{MC@NLO+Herwig} and \textsc{Powheg+Pythia} event generators for $30 \GeV < p_{\mathrm{T}} < 40 \GeV$. The uncertainties shown include statistical and systematic sources, added in quadrature.}
\label{fig:diff1}
\end{figure}
\begin{table}[!h]
\vspace{0.5cm}
\caption{Unfolded values for $\langle\rho(r)\rangle$, together with statistical and systematic uncertainties for $30 \GeV < p_{\mathrm{T}} < 40 \GeV$.}
\normalsize
\label{tabDiff1}
\begin{center}
\begin{tabular}{ccc}
\hline
$r$ & $\langle\rho_b (r)\rangle$ [$b$-jets] & $\langle\rho_{\mathrm{l}} (r)\rangle$ [light jets]\\
\hline
0.02 & 3.84  $\pm$  0.15 $^{+ 0.29 }_{- 0.36 }$ & 7.64  $\pm$  0.27 $^{+ 0.93 }_{- 1.10 }$ \\[3pt] 
0.06 & 6.06  $\pm$  0.14 $^{+ 0.31 }_{- 0.36 }$ & 6.10  $\pm$  0.16 $^{+ 0.48 }_{- 0.47 }$ \\[3pt] 
0.10 & 5.20  $\pm$  0.11 $^{+ 0.24 }_{- 0.23 }$ & 3.75  $\pm$  0.10 $^{+ 0.32 }_{- 0.33 }$ \\[3pt] 
0.14 & 3.45  $\pm$  0.09 $^{+ 0.12 }_{- 0.13 }$ & 2.28  $\pm$  0.07 $^{+ 0.14 }_{- 0.16 }$ \\[3pt] 
0.18 & 2.21  $\pm$  0.06 $^{+ 0.13 }_{- 0.11 }$ & 1.50  $\pm$  0.05 $^{+ 0.14 }_{- 0.12 }$ \\[3pt] 
0.22 & 1.58  $\pm$  0.04 $^{+ 0.10 }_{- 0.11 }$ & 1.08  $\pm$  0.03 $^{+ 0.09 }_{- 0.10 }$ \\[3pt] 
0.26 & 1.15  $\pm$  0.03 $^{+ 0.13 }_{- 0.13 }$ & 0.83  $\pm$  0.03 $^{+ 0.11 }_{- 0.09 }$ \\[3pt] 
0.30 & 0.80  $\pm$  0.02 $^{+ 0.08 }_{- 0.07 }$ & 0.64  $\pm$  0.02 $^{+ 0.07 }_{- 0.08 }$ \\[3pt] 
0.34 & 0.60  $\pm$  0.01 $^{+ 0.06 }_{- 0.06 }$ & 0.53  $\pm$  0.01 $^{+ 0.07 }_{- 0.08 }$ \\[3pt] 
0.38 & 0.32  $\pm$  0.01 $^{+ 0.04 }_{- 0.04 }$ & 0.28  $\pm$  0.01 $^{+ 0.04 }_{- 0.04 }$ \\[3pt] 
\hline
\end{tabular}
\end{center}
\end{table}

\begin{figure}[!h]
\vspace{0.8cm}
\includegraphics[width=8.5cm,height=10.5cm]{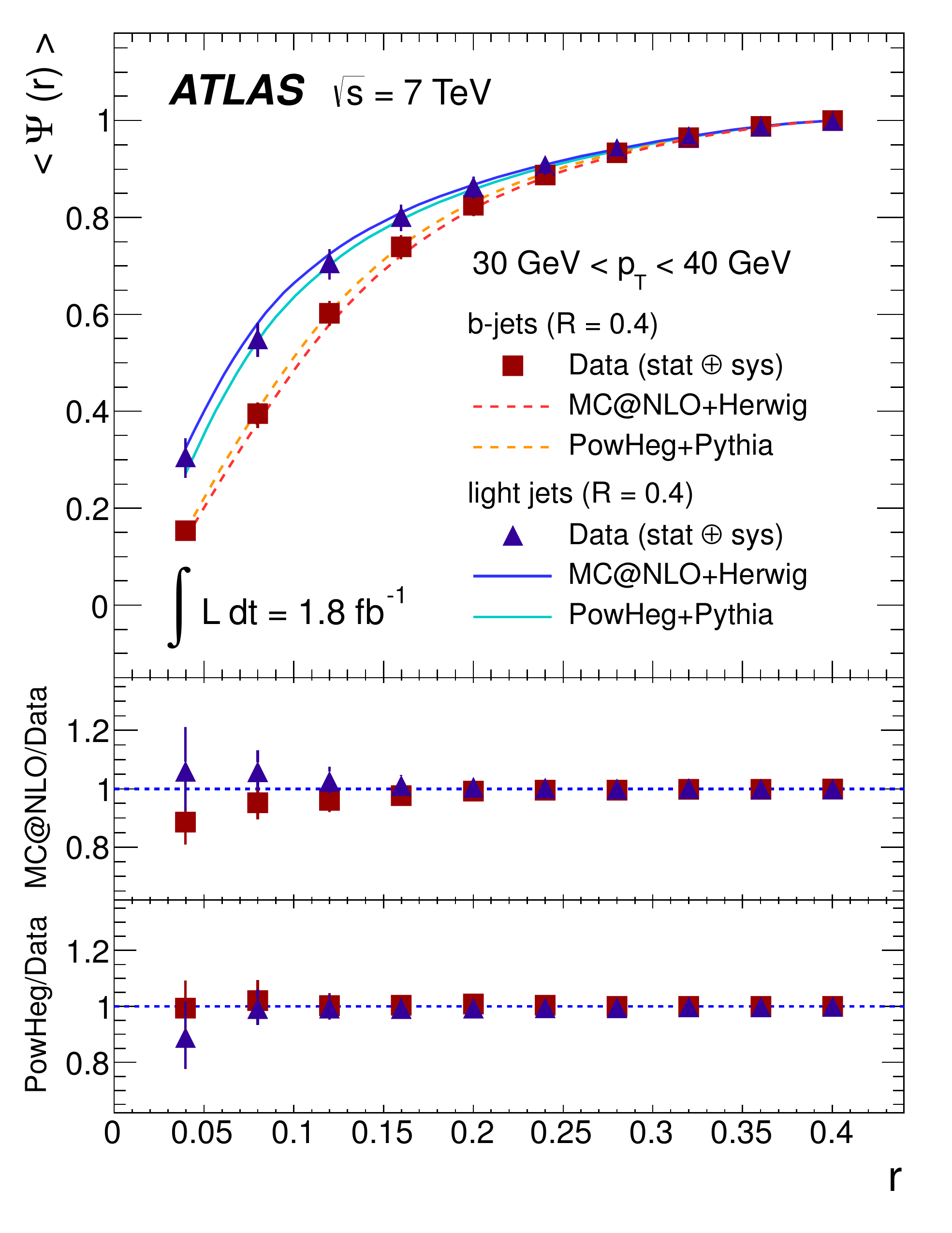}
\caption{Integrated jet shapes $\langle\Psi(r)\rangle$ as a function of the radius $r$ for light jets (triangles) and $b$-jets (squares). The data are compared to \textsc{MC@NLO+Herwig} and \textsc{Powheg+Pythia} event generators for $30 \GeV < p_{\mathrm{T}} < 40 \GeV$. The uncertainties shown include statistical and systematic sources, added in quadrature.}
\label{fig:int1}
\end{figure}
\begin{table}[!h]
\vspace{0.81cm}
\caption{Unfolded values for $\langle\Psi(r)\rangle$, together with statistical and systematic uncertainties for $30 \GeV < p_{\mathrm{T}} < 40 \GeV$.}
\normalsize
\label{tabInt1}
\begin{center}
\begin{tabular}{ccc}
\hline
$r$ & $\langle\Psi_b (r)\rangle$ [$b$-jets] & $\langle\Psi_{\mathrm{l}} (r)\rangle$ [light jets]\\
\hline
0.04 & 0.154  $\pm$  0.006 $^{+ 0.012 }_{- 0.014 }$ & 0.306  $\pm$  0.011 $^{+ 0.037 }_{- 0.043 }$ \\[3pt]
0.08 & 0.395  $\pm$  0.007 $^{+ 0.023 }_{- 0.028 }$ & 0.550  $\pm$  0.009 $^{+ 0.031 }_{- 0.037 }$ \\[3pt]
0.12 & 0.602  $\pm$  0.006 $^{+ 0.025 }_{- 0.026 }$ & 0.706  $\pm$  0.007 $^{+ 0.028 }_{- 0.034 }$ \\[3pt]
0.16 & 0.739  $\pm$  0.004 $^{+ 0.025 }_{- 0.025 }$ & 0.802  $\pm$  0.005 $^{+ 0.025 }_{- 0.030 }$ \\[3pt]
0.20 & 0.825  $\pm$  0.003 $^{+ 0.020 }_{- 0.023 }$ & 0.863  $\pm$  0.004 $^{+ 0.020 }_{- 0.025 }$ \\[3pt]
0.24 & 0.887  $\pm$  0.003 $^{+ 0.016 }_{- 0.017 }$ & 0.907  $\pm$  0.003 $^{+ 0.016 }_{- 0.019 }$ \\[3pt]
0.28 & 0.934  $\pm$  0.002 $^{+ 0.012 }_{- 0.012 }$ & 0.942  $\pm$  0.002 $^{+ 0.011 }_{- 0.014 }$ \\[3pt]
0.32 & 0.964  $\pm$  0.001 $^{+ 0.007 }_{- 0.007 }$ & 0.967  $\pm$  0.001 $^{+ 0.007 }_{- 0.008 }$ \\[3pt]
0.36 & 0.988  $\pm$  0.001 $^{+ 0.004 }_{- 0.002 }$ & 0.989  $\pm$  0.001 $^{+ 0.003 }_{- 0.003 }$ \\[3pt]
0.40 & 1.000  & 1.000 \\[3pt]
\hline
\end{tabular}
\end{center}
\end{table}

\begin{figure}[!h]
\vspace{0.8cm}
\includegraphics[width=8.5cm,height=10.5cm]{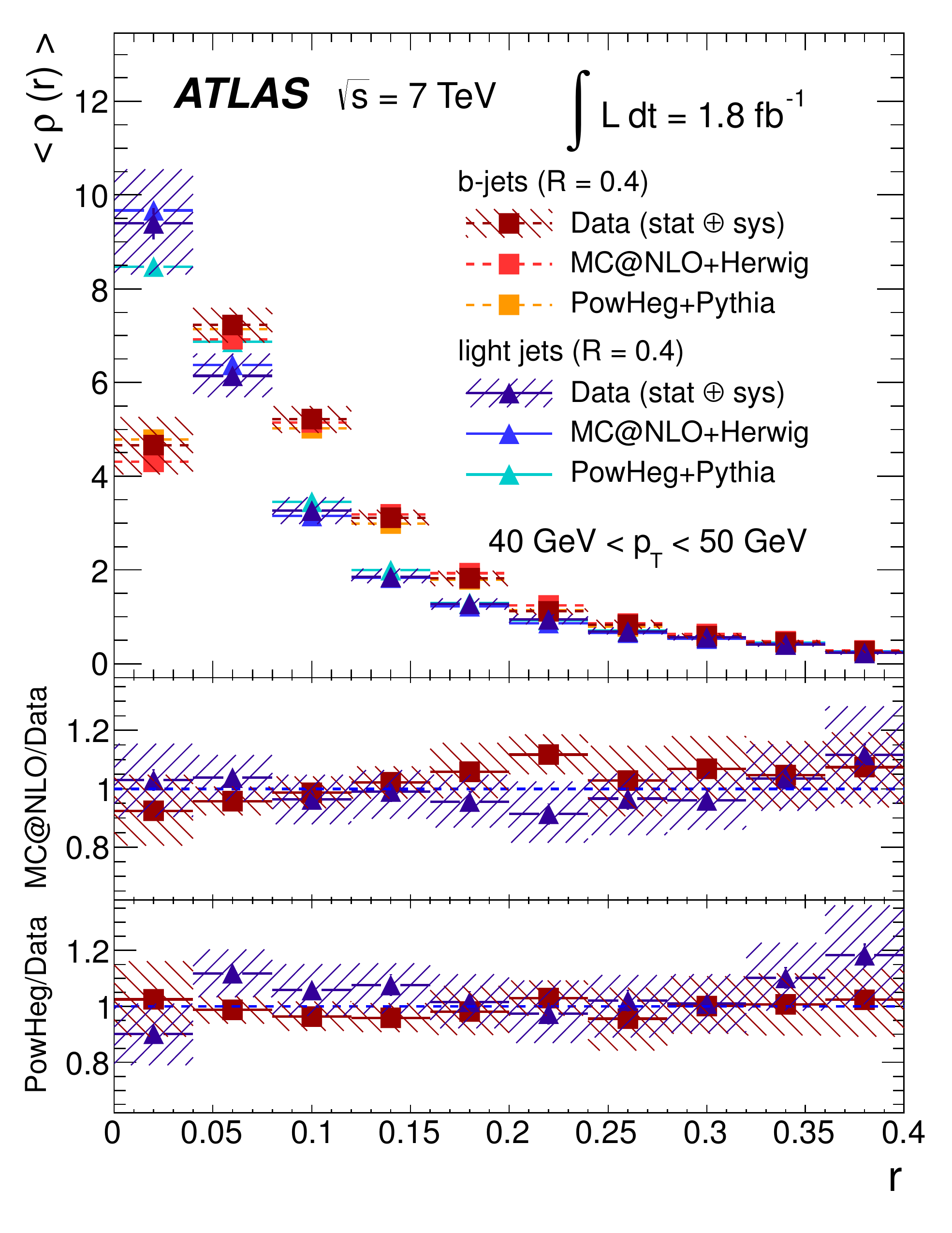}
\caption{Differential jet shapes $\langle\rho(r)\rangle$ as a function of the radius $r$ for light jets (triangles) and $b$-jets (squares). The data are compared to \textsc{MC@NLO+Herwig} and \textsc{Powheg+Pythia} event generators for $40 \GeV < p_{\mathrm{T}} < 50 \GeV$. The uncertainties shown include statistical and systematic sources, added in quadrature.}
\label{fig:diff2}
\end{figure}

\begin{table}[!h]
\vspace{0.8cm}
\caption{Unfolded values for $\langle\rho(r)\rangle$, together with statistical and systematic uncertainties for $40 \GeV < p_{\mathrm{T}} < 50 \GeV$.}
\normalsize
\label{tabDiff2}
\begin{center}
\begin{tabular}{ccc}
\hline
$r$ & $\langle\rho_b (r)\rangle$ [$b$-jets] & $\langle\rho_{\mathrm{l}} (r)\rangle$ [light jets]\\
\hline
0.02 & 4.66  $\pm$  0.15 $^{+ 0.58 }_{- 0.61 }$ & 9.39  $\pm$  0.34 $^{+ 1.10 }_{- 1.10 }$ \\[3pt] 
0.06 & 7.23  $\pm$  0.14 $^{+ 0.33 }_{- 0.35 }$ & 6.14  $\pm$  0.17 $^{+ 0.44 }_{- 0.43 }$ \\[3pt] 
0.10 & 5.22  $\pm$  0.11 $^{+ 0.25 }_{- 0.28 }$ & 3.27  $\pm$  0.10 $^{+ 0.27 }_{- 0.27 }$ \\[3pt] 
0.14 & 3.12  $\pm$  0.07 $^{+ 0.15 }_{- 0.15 }$ & 1.85  $\pm$  0.07 $^{+ 0.16 }_{- 0.12 }$ \\[3pt] 
0.18 & 1.83  $\pm$  0.05 $^{+ 0.15 }_{- 0.17 }$ & 1.28  $\pm$  0.05 $^{+ 0.11 }_{- 0.11 }$ \\[3pt] 
0.22 & 1.12  $\pm$  0.03 $^{+ 0.06 }_{- 0.06 }$ & 0.95  $\pm$  0.04 $^{+ 0.10 }_{- 0.11 }$ \\[3pt] 
0.26 & 0.83  $\pm$  0.02 $^{+ 0.10 }_{- 0.09 }$ & 0.69  $\pm$  0.03 $^{+ 0.08 }_{- 0.05 }$ \\[3pt] 
0.30 & 0.59  $\pm$  0.02 $^{+ 0.06 }_{- 0.06 }$ & 0.56  $\pm$  0.02 $^{+ 0.05 }_{- 0.05 }$ \\[3pt] 
0.34 & 0.46  $\pm$  0.01 $^{+ 0.05 }_{- 0.05 }$ & 0.41  $\pm$  0.01 $^{+ 0.04 }_{- 0.04 }$ \\[3pt] 
0.38 & 0.26  $\pm$  0.01 $^{+ 0.03 }_{- 0.03 }$ & 0.23  $\pm$  0.01 $^{+ 0.03 }_{- 0.03 }$ \\[3pt] 
\hline
\end{tabular}
\end{center}
\end{table}

\begin{figure}[!h]
\vspace{0.8cm}
\includegraphics[width=8.5cm,height=10.5cm]{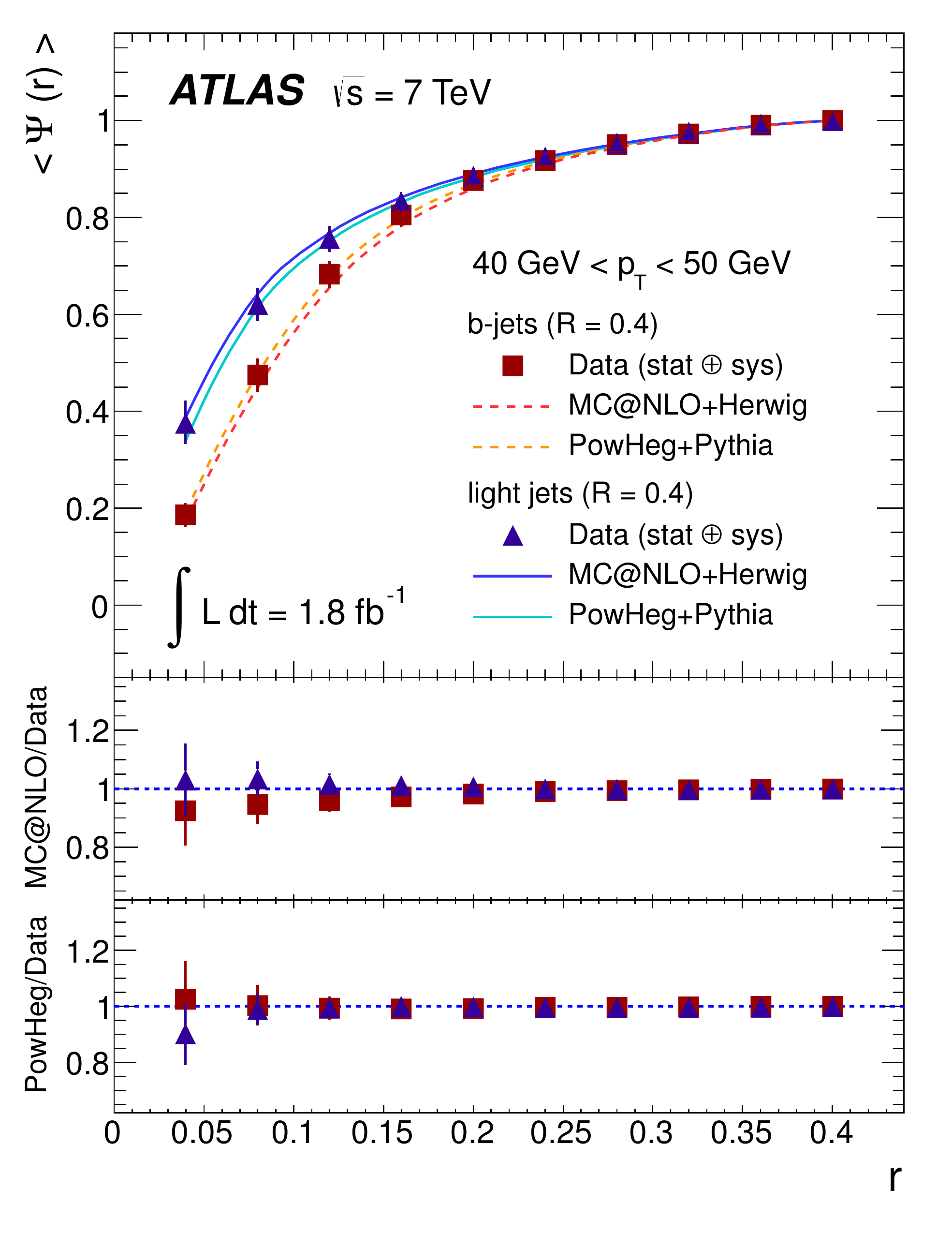}
\caption{Integrated jet shapes $\langle\Psi(r)\rangle$ as a function of the radius $r$ for light jets (triangles) and $b$-jets (squares). The data are compared to \textsc{MC@NLO+Herwig} and \textsc{Powheg+Pythia} event generators for $40 \GeV < p_{\mathrm{T}} < 50 \GeV$. The uncertainties shown include statistical and systematic sources, added in quadrature.}
\label{fig:int2}
\end{figure}

\begin{table}[!h]
\vspace{0.81cm}
\caption{Unfolded values for $\langle\Psi(r)\rangle$, together with statistical and systematic uncertainties for $40 \GeV < p_{\mathrm{T}} < 50 \GeV$.}
\normalsize
\label{tabInt2}
\begin{center}
\begin{tabular}{ccc}
\hline
$r$ & $\langle\Psi_b (r)\rangle$ [$b$-jets] & $\langle\Psi_{\mathrm{l}} (r)\rangle$ [light jets]\\
\hline
0.04 & 0.187  $\pm$  0.006 $^{+ 0.023 }_{- 0.024 }$ & 0.376  $\pm$  0.013 $^{+ 0.044 }_{- 0.043 }$ \\[3pt]
0.08 & 0.475  $\pm$  0.007 $^{+ 0.033 }_{- 0.034 }$ & 0.621  $\pm$  0.011 $^{+ 0.032 }_{- 0.034 }$ \\[3pt]
0.12 & 0.683  $\pm$  0.005 $^{+ 0.027 }_{- 0.029 }$ & 0.757  $\pm$  0.008 $^{+ 0.025 }_{- 0.027 }$ \\[3pt]
0.16 & 0.805  $\pm$  0.004 $^{+ 0.023 }_{- 0.025 }$ & 0.832  $\pm$  0.006 $^{+ 0.021 }_{- 0.022 }$ \\[3pt]
0.20 & 0.876  $\pm$  0.003 $^{+ 0.017 }_{- 0.018 }$ & 0.885  $\pm$  0.004 $^{+ 0.017 }_{- 0.018 }$ \\[3pt]
0.24 & 0.918  $\pm$  0.002 $^{+ 0.015 }_{- 0.016 }$ & 0.925  $\pm$  0.003 $^{+ 0.012 }_{- 0.014 }$ \\[3pt]
0.28 & 0.950  $\pm$  0.002 $^{+ 0.010 }_{- 0.011 }$ & 0.953  $\pm$  0.002 $^{+ 0.010 }_{- 0.011 }$ \\[3pt]
0.32 & 0.973  $\pm$  0.001 $^{+ 0.007 }_{- 0.006 }$ & 0.976  $\pm$  0.001 $^{+ 0.006 }_{- 0.006 }$ \\[3pt]
0.36 & 0.990  $\pm$  0.001 $^{+ 0.003 }_{- 0.002 }$ & 0.992  $\pm$  0.001 $^{+ 0.003 }_{- 0.003 }$ \\[3pt]
0.40 & 1.000 & 1.000 \\[3pt]
\hline
\end{tabular}
\end{center}
\end{table}

\begin{figure}[!h]
\vspace{0.8cm}
\includegraphics[width=8.5cm,height=10.5cm]{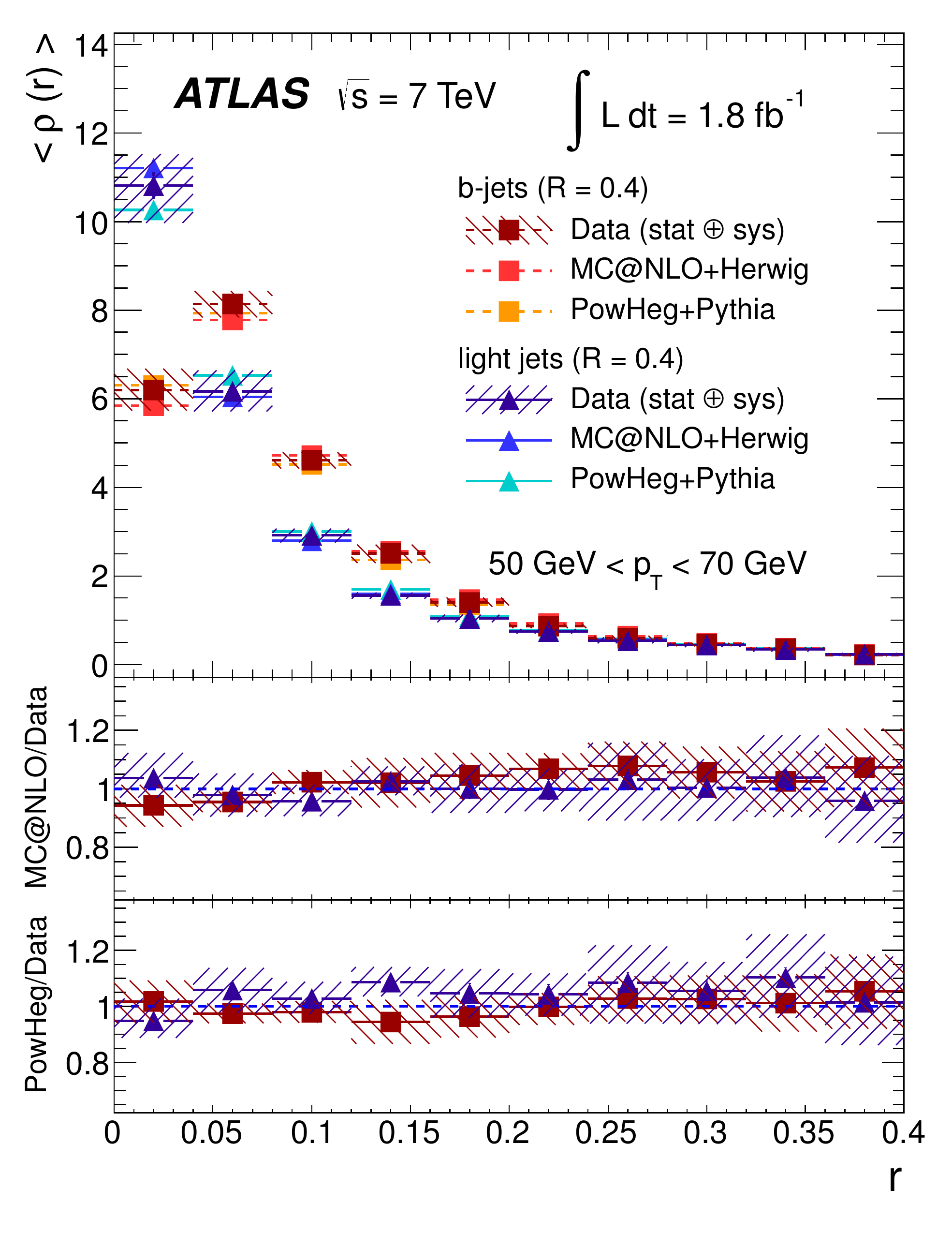}
\caption{Differential jet shapes $\langle\rho(r)\rangle$ as a function of the radius $r$ for light jets (triangles) and $b$-jets (squares). The data are compared to \textsc{MC@NLO+Herwig} and \textsc{Powheg+Pythia} event generators for $50 \GeV < p_{\mathrm{T}} < 70 \GeV$. The uncertainties shown include statistical and systematic sources, added in quadrature.}
\label{fig:diff3}
\end{figure}

\begin{table}[!h]
\vspace{0.8cm}
\caption{Unfolded values for $\langle\rho(r)\rangle$, together with statistical and systematic uncertainties for $50 \GeV < p_{\mathrm{T}} < 70 \GeV$.}
\normalsize
\label{tabDiff3}
\begin{center}
\begin{tabular}{ccc}
\hline
$r$ & $\langle\rho_b (r)\rangle$ [$b$-jets] & $\langle\rho_{\mathrm{l}} (r)\rangle$ [light jets]\\
\hline
0.02 & 6.19  $\pm$  0.13 $^{+ 0.46 }_{- 0.44 }$ & 10.82  $\pm$  0.31 $^{+ 0.64 }_{- 0.84 }$ \\[3pt] 
0.06 & 8.14  $\pm$  0.11 $^{+ 0.27 }_{- 0.29 }$ & 6.17  $\pm$  0.14 $^{+ 0.45 }_{- 0.44 }$ \\[3pt] 
0.10 & 4.62  $\pm$  0.06 $^{+ 0.17 }_{- 0.18 }$ & 2.92  $\pm$  0.08 $^{+ 0.14 }_{- 0.15 }$ \\[3pt] 
0.14 & 2.50  $\pm$  0.04 $^{+ 0.20 }_{- 0.21 }$ & 1.56  $\pm$  0.05 $^{+ 0.05 }_{- 0.06 }$ \\[3pt] 
0.18 & 1.40  $\pm$  0.03 $^{+ 0.11 }_{- 0.10 }$ & 1.04  $\pm$  0.04 $^{+ 0.08 }_{- 0.08 }$ \\[3pt] 
0.22 & 0.87  $\pm$  0.02 $^{+ 0.05 }_{- 0.04 }$ & 0.75  $\pm$  0.03 $^{+ 0.05 }_{- 0.05 }$ \\[3pt] 
0.26 & 0.60  $\pm$  0.01 $^{+ 0.05 }_{- 0.04 }$ & 0.54  $\pm$  0.02 $^{+ 0.07 }_{- 0.06 }$ \\[3pt] 
0.30 & 0.45  $\pm$  0.01 $^{+ 0.04 }_{- 0.04 }$ & 0.44  $\pm$  0.01 $^{+ 0.05 }_{- 0.04 }$ \\[3pt] 
0.34 & 0.36  $\pm$  0.01 $^{+ 0.04 }_{- 0.04 }$ & 0.34  $\pm$  0.01 $^{+ 0.04 }_{- 0.05 }$ \\[3pt] 
0.38 & 0.21  $\pm$  0.00 $^{+ 0.03 }_{- 0.03 }$ & 0.23  $\pm$  0.01 $^{+ 0.03 }_{- 0.04 }$ \\[3pt] 
\hline
\end{tabular}
\end{center}
\end{table}

\begin{figure}[!h]
\vspace{0.8cm}
\includegraphics[width=8.5cm,height=10.5cm]{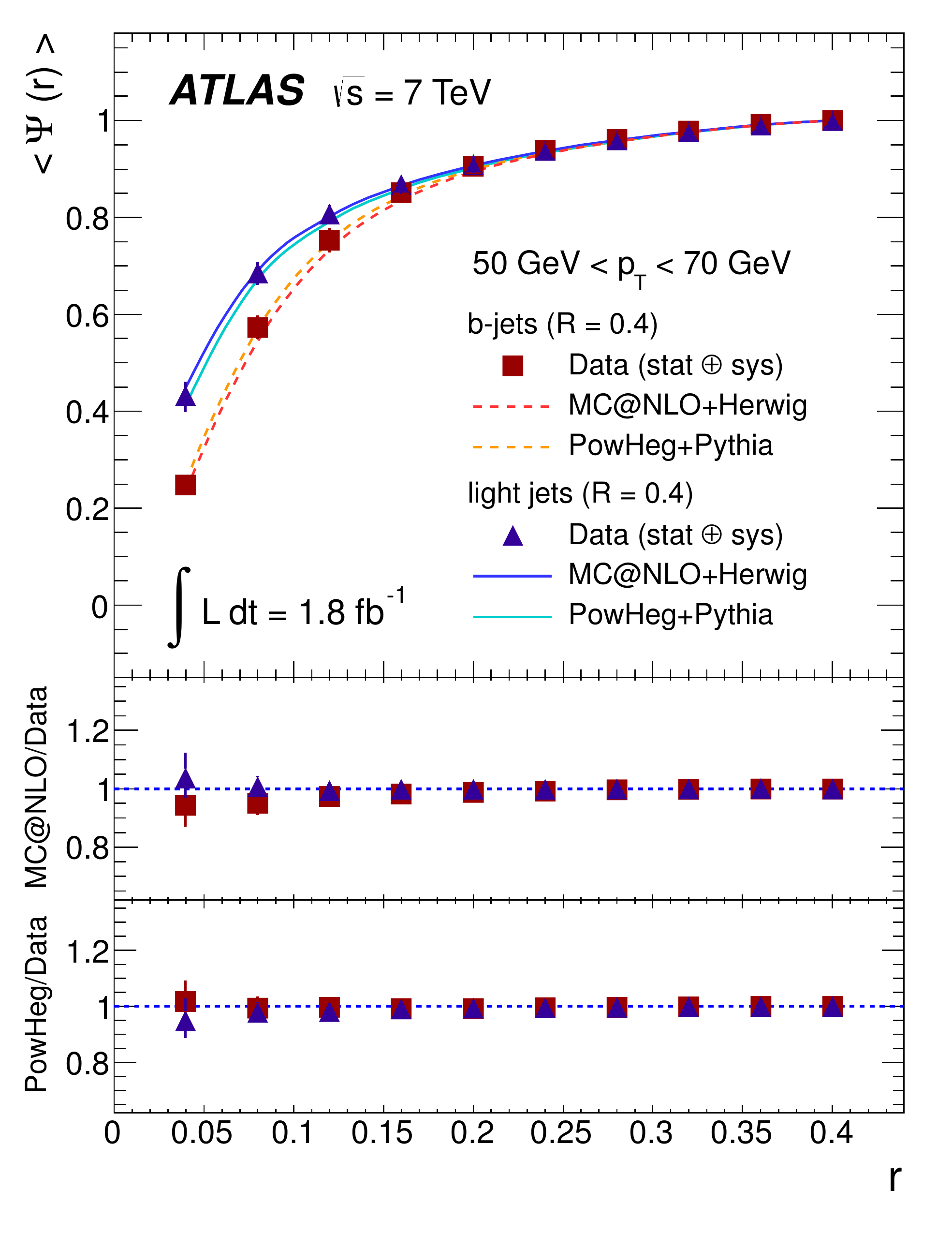}
\caption{Integrated jet shapes $\langle\Psi(r)\rangle$ as a function of the radius $r$ for light jets (triangles) and $b$-jets (squares). The data are compared to \textsc{MC@NLO+Herwig} and \textsc{Powheg+Pythia} event generators for $50 \GeV < p_{\mathrm{T}} < 70 \GeV$. The uncertainties shown include statistical and systematic sources, added in quadrature.}
\label{fig:int3}
\end{figure}

\begin{table}[!h]
\vspace{0.81cm}
\caption{Unfolded values for $\langle\Psi(r)\rangle$, together with statistical and systematic uncertainties for $50 \GeV < p_{\mathrm{T}} < 70 \GeV$.}
\normalsize
\label{tabInt3}
\begin{center}
\begin{tabular}{ccc}
\hline
$r$ & $\langle\Psi_b (r)\rangle$ [$b$-jets] & $\langle\Psi_{\mathrm{l}} (r)\rangle$ [light jets]\\
\hline
0.04 & 0.248  $\pm$  0.005 $^{+ 0.019 }_{- 0.018 }$ & 0.433  $\pm$  0.012 $^{+ 0.026 }_{- 0.034 }$ \\[3pt]
0.08 & 0.573  $\pm$  0.005 $^{+ 0.024 }_{- 0.023 }$ & 0.686  $\pm$  0.009 $^{+ 0.020 }_{- 0.024 }$ \\[3pt]
0.12 & 0.753  $\pm$  0.004 $^{+ 0.025 }_{- 0.025 }$ & 0.807  $\pm$  0.006 $^{+ 0.017 }_{- 0.019 }$ \\[3pt]
0.16 & 0.851  $\pm$  0.003 $^{+ 0.019 }_{- 0.018 }$ & 0.868  $\pm$  0.004 $^{+ 0.017 }_{- 0.019 }$ \\[3pt]
0.20 & 0.905  $\pm$  0.002 $^{+ 0.015 }_{- 0.015 }$ & 0.909  $\pm$  0.003 $^{+ 0.014 }_{- 0.016 }$ \\[3pt]
0.24 & 0.938  $\pm$  0.001 $^{+ 0.012 }_{- 0.013 }$ & 0.939  $\pm$  0.002 $^{+ 0.012 }_{- 0.014 }$ \\[3pt]
0.28 & 0.961  $\pm$  0.001 $^{+ 0.008 }_{- 0.009 }$ & 0.960  $\pm$  0.002 $^{+ 0.008 }_{- 0.009 }$ \\[3pt]
0.32 & 0.978  $\pm$  0.001 $^{+ 0.005 }_{- 0.005 }$ & 0.977  $\pm$  0.001 $^{+ 0.006 }_{- 0.006 }$ \\[3pt]
0.36 & 0.992  $\pm$  0.000 $^{+ 0.003 }_{- 0.002 }$ & 0.990  $\pm$  0.001 $^{+ 0.003 }_{- 0.003 }$ \\[3pt]
0.40 & 1.000 & 1.000 \\[3pt]
\hline
\end{tabular}
\end{center}
\end{table}

\begin{figure}[!h]
\vspace{0.8cm}
\includegraphics[width=8.5cm,height=10.5cm]{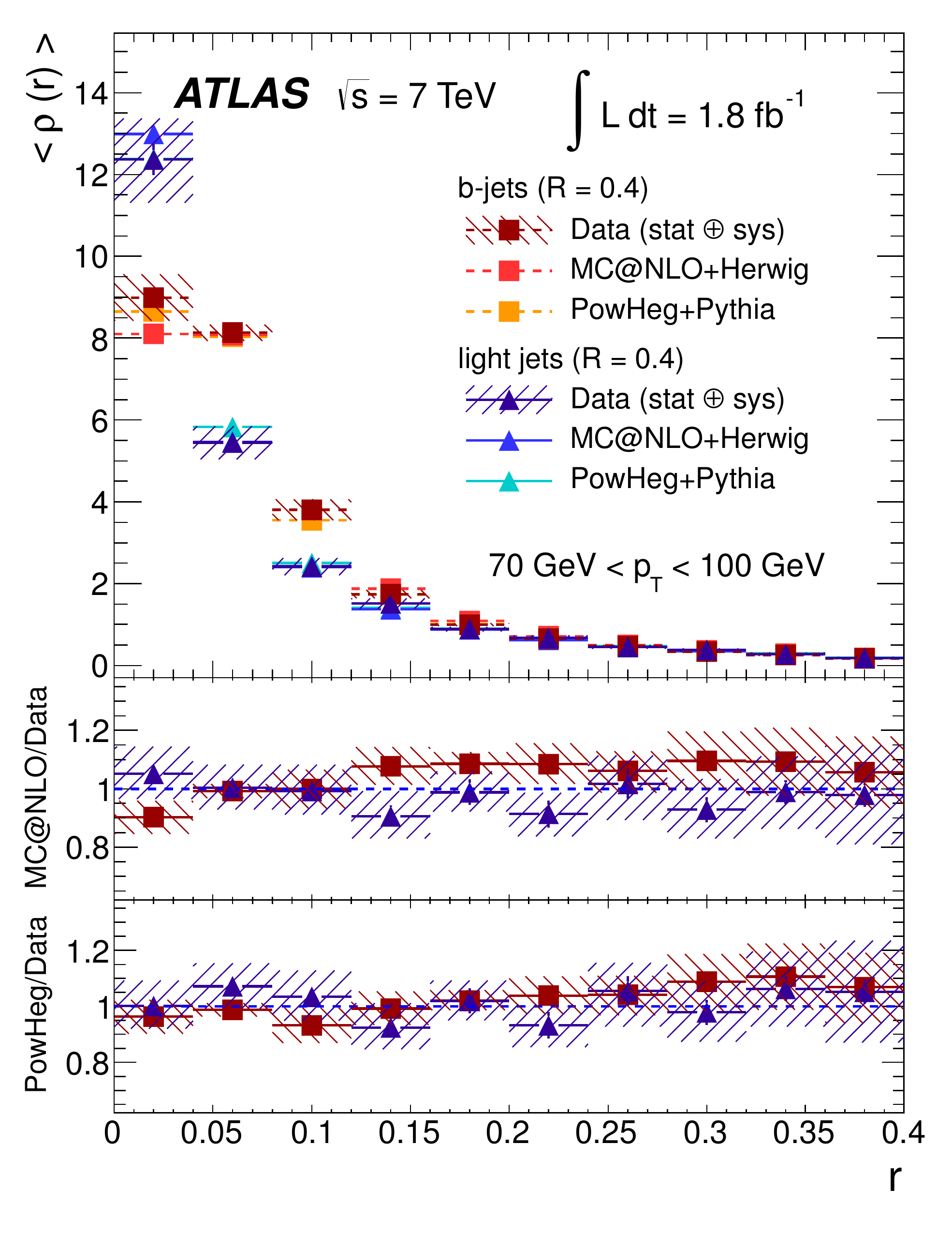}
\caption{Differential jet shapes $\langle\rho(r)\rangle$ as a function of the radius $r$ for light jets (triangles) and $b$-jets (squares). The data are compared to \textsc{MC@NLO+Herwig} and \textsc{Powheg+Pythia} event generators for $70 \GeV < p_{\mathrm{T}} < 100 \GeV$. The uncertainties shown include statistical and systematic sources, added in quadrature.}
\label{fig:diff4}
\end{figure}

\begin{table}[!h]
\vspace{0.8cm}
\caption{Unfolded values for $\langle\rho(r)\rangle$, together with statistical and systematic uncertainties for $70 \GeV < p_{\mathrm{T}} < 100 \GeV$.}
\normalsize
\label{tabDiff4}
\begin{center}
\begin{tabular}{ccc}
\hline
$r$ & $\langle\rho_b (r)\rangle$ [$b$-jets] & $\langle\rho_{\mathrm{l}} (r)\rangle$ [light jets]\\
\hline
0.02 & 8.98  $\pm$  0.15 $^{+ 0.55 }_{- 0.54 }$ & 12.37  $\pm$  0.38 $^{+ 0.93 }_{- 1.10 }$ \\[3pt] 
0.06 & 8.14  $\pm$  0.10 $^{+ 0.17 }_{- 0.17 }$ & 5.44  $\pm$  0.16 $^{+ 0.38 }_{- 0.39 }$ \\[3pt] 
0.10 & 3.80  $\pm$  0.05 $^{+ 0.25 }_{- 0.25 }$ & 2.42  $\pm$  0.08 $^{+ 0.18 }_{- 0.21 }$ \\[3pt] 
0.14 & 1.74  $\pm$  0.03 $^{+ 0.10 }_{- 0.10 }$ & 1.52  $\pm$  0.06 $^{+ 0.11 }_{- 0.13 }$ \\[3pt] 
0.18 & 1.00  $\pm$  0.02 $^{+ 0.03 }_{- 0.03 }$ & 0.89  $\pm$  0.04 $^{+ 0.05 }_{- 0.05 }$ \\[3pt] 
0.22 & 0.66  $\pm$  0.01 $^{+ 0.04 }_{- 0.04 }$ & 0.68  $\pm$  0.03 $^{+ 0.05 }_{- 0.04 }$ \\[3pt] 
0.26 & 0.47  $\pm$  0.01 $^{+ 0.03 }_{- 0.03 }$ & 0.45  $\pm$  0.02 $^{+ 0.05 }_{- 0.04 }$ \\[3pt] 
0.30 & 0.34  $\pm$  0.01 $^{+ 0.03 }_{- 0.03 }$ & 0.38  $\pm$  0.02 $^{+ 0.04 }_{- 0.04 }$ \\[3pt] 
0.34 & 0.26  $\pm$  0.01 $^{+ 0.03 }_{- 0.03 }$ & 0.28  $\pm$  0.01 $^{+ 0.03 }_{- 0.03 }$ \\[3pt] 
0.38 & 0.17  $\pm$  0.00 $^{+ 0.02 }_{- 0.02 }$ & 0.18  $\pm$  0.01 $^{+ 0.03 }_{- 0.03 }$ \\[3pt] 
\hline
\end{tabular}
\end{center}
\end{table}

\begin{figure}[!h]
\vspace{0.8cm}
\includegraphics[width=8.5cm,height=10.5cm]{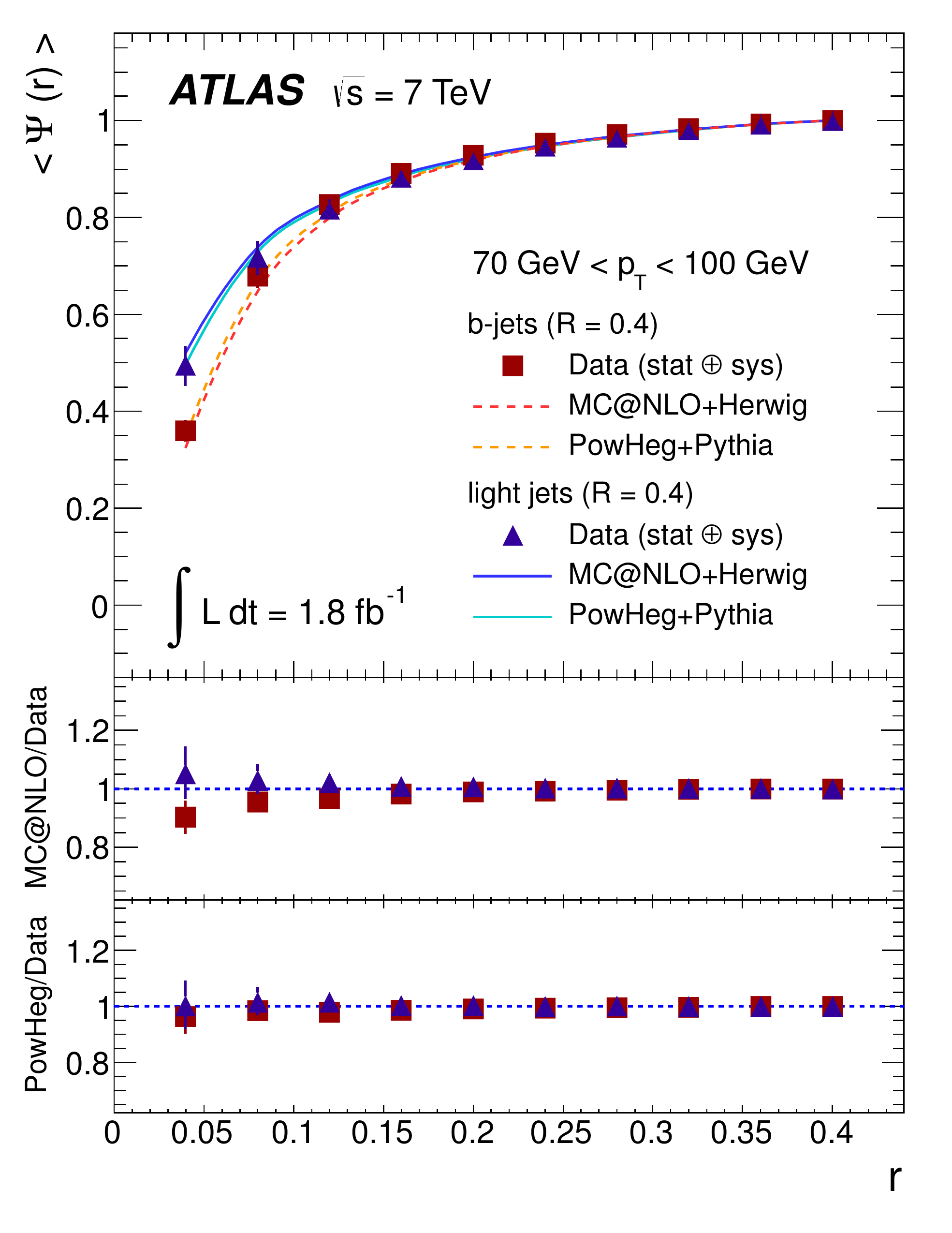}
\caption{Integrated jet shapes $\langle\Psi(r)\rangle$ as a function of the radius $r$ for light jets (triangles) and $b$-jets (squares). The data are compared to \textsc{MC@NLO+Herwig} and \textsc{Powheg+Pythia} event generators for $70 \GeV < p_{\mathrm{T}} < 100 \GeV$. The uncertainties shown include statistical and systematic sources, added in quadrature.}
\label{fig:int4}
\vspace{0.8cm}
\end{figure}

\begin{table}[!h]
\caption{Unfolded values for $\langle\Psi(r)\rangle$, together with statistical and systematic uncertainties for $70 \GeV < p_{\mathrm{T}} < 100 \GeV$.}
\normalsize
\label{tabInt4}
\begin{center}
\begin{tabular}{ccc}
\hline
$r$ & $\langle\Psi_b (r)\rangle$ [$b$-jets] & $\langle\Psi_{\mathrm{l}} (r)\rangle$ [light jets]\\
\hline
0.04 & 0.359  $\pm$  0.006 $^{+ 0.022 }_{- 0.021 }$ & 0.495  $\pm$  0.015 $^{+ 0.037 }_{- 0.042 }$ \\[3pt]
0.08 & 0.678  $\pm$  0.005 $^{+ 0.023 }_{- 0.023 }$ & 0.718  $\pm$  0.010 $^{+ 0.032 }_{- 0.037 }$ \\[3pt]
0.12 & 0.827  $\pm$  0.003 $^{+ 0.017 }_{- 0.018 }$ & 0.818  $\pm$  0.007 $^{+ 0.019 }_{- 0.021 }$ \\[3pt]
0.16 & 0.891  $\pm$  0.002 $^{+ 0.012 }_{- 0.013 }$ & 0.883  $\pm$  0.005 $^{+ 0.012 }_{- 0.014 }$ \\[3pt]
0.20 & 0.928  $\pm$  0.002 $^{+ 0.011 }_{- 0.012 }$ & 0.919  $\pm$  0.004 $^{+ 0.010 }_{- 0.011 }$ \\[3pt]
0.24 & 0.954  $\pm$  0.001 $^{+ 0.009 }_{- 0.009 }$ & 0.947  $\pm$  0.003 $^{+ 0.008 }_{- 0.009 }$ \\[3pt]
0.28 & 0.972  $\pm$  0.001 $^{+ 0.006 }_{- 0.007 }$ & 0.965  $\pm$  0.002 $^{+ 0.007 }_{- 0.008 }$ \\[3pt]
0.32 & 0.984  $\pm$  0.001 $^{+ 0.004 }_{- 0.004 }$ & 0.981  $\pm$  0.001 $^{+ 0.004 }_{- 0.005 }$ \\[3pt]
0.36 & 0.993  $\pm$  0.000 $^{+ 0.002 }_{- 0.002 }$ & 0.992  $\pm$  0.001 $^{+ 0.002 }_{- 0.002 }$ \\[3pt]
0.40 & 1.000 & 1.000 \\[3pt]
\hline
\end{tabular}
\end{center}
\end{table}

\begin{figure}[!h]
\vspace{0.8cm}
\includegraphics[width=8.5cm,height=10.5cm]{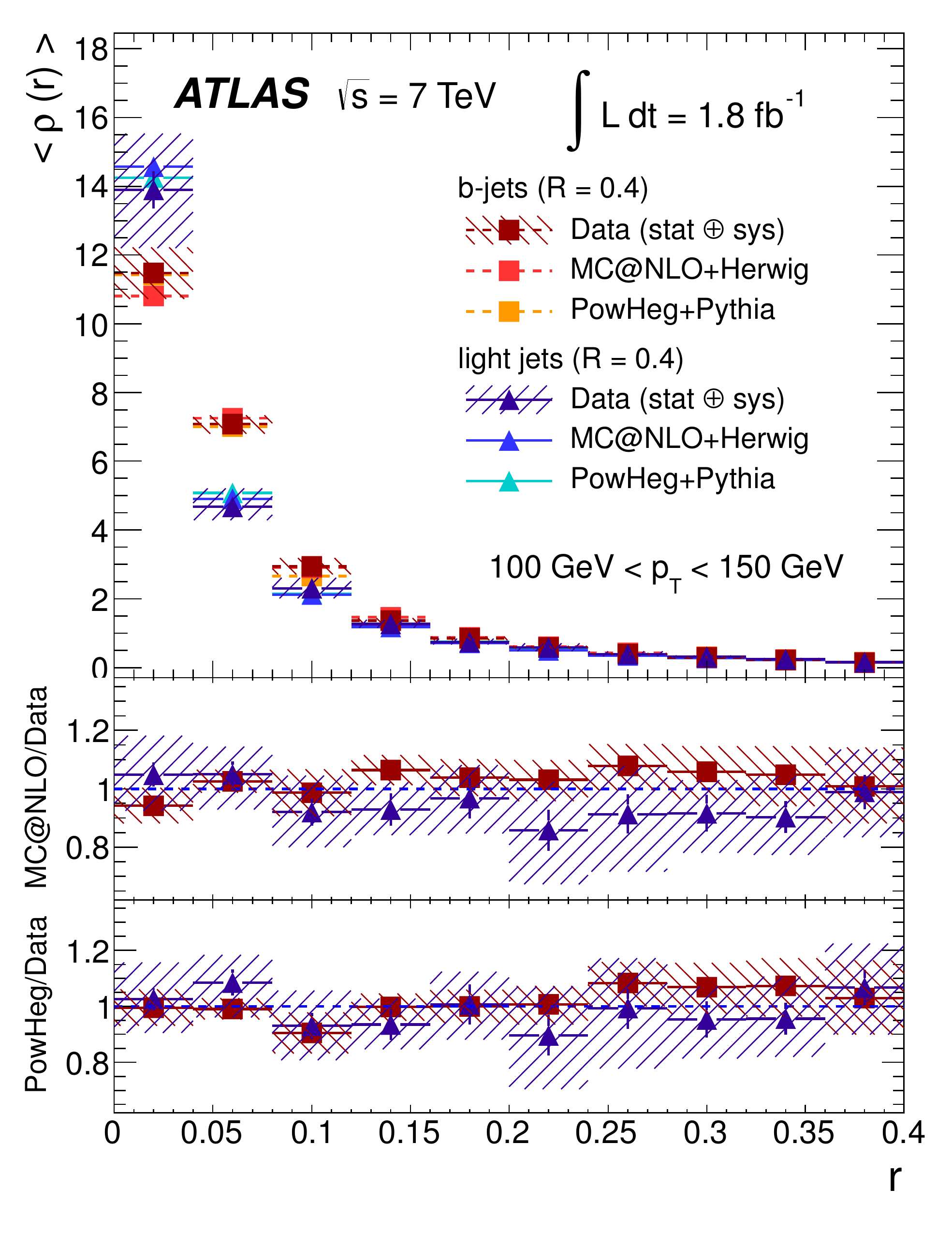}
\caption{Differential jet shapes $\langle\rho(r)\rangle$ as a function of the radius $r$ for light jets (triangles) and $b$-jets (squares). The data are compared to \textsc{MC@NLO+Herwig} and \textsc{Powheg+Pythia} event generators for $100 \GeV < p_{\mathrm{T}} < 150 \GeV$. The uncertainties shown include statistical and systematic sources, added in quadrature.}
\label{fig:diff5}
\end{figure}

\begin{table}[!h]
\vspace{0.8cm}
\caption{Unfolded values for $\langle\rho(r)\rangle$, together with statistical and systematic uncertainties for $100 \GeV < p_{\mathrm{T}} < 150 \GeV$.}
\normalsize
\label{tabDiff5}
\begin{center}
\begin{tabular}{ccc}
\hline
$r$ & $\langle\rho_b (r)\rangle$ [$b$-jets] & $\langle\rho_{\mathrm{l}} (r)\rangle$ [light jets]\\
\hline
0.02 & 11.48  $\pm$  0.20 $^{+ 0.71 }_{- 0.74 }$ & 13.89  $\pm$  0.54 $^{+ 1.60 }_{- 1.70 }$ \\[3pt] 
0.06 & 7.08  $\pm$  0.11 $^{+ 0.24 }_{- 0.25 }$ & 4.68  $\pm$  0.20 $^{+ 0.50 }_{- 0.37 }$ \\[3pt] 
0.10 & 2.94  $\pm$  0.05 $^{+ 0.23 }_{- 0.23 }$ & 2.31  $\pm$  0.11 $^{+ 0.28 }_{- 0.29 }$ \\[3pt] 
0.14 & 1.37  $\pm$  0.03 $^{+ 0.06 }_{- 0.06 }$ & 1.27  $\pm$  0.07 $^{+ 0.09 }_{- 0.10 }$ \\[3pt] 
0.18 & 0.85  $\pm$  0.02 $^{+ 0.05 }_{- 0.05 }$ & 0.74  $\pm$  0.05 $^{+ 0.08 }_{- 0.07 }$ \\[3pt] 
0.22 & 0.58  $\pm$  0.02 $^{+ 0.04 }_{- 0.03 }$ & 0.58  $\pm$  0.05 $^{+ 0.12 }_{- 0.10 }$ \\[3pt] 
0.26 & 0.39  $\pm$  0.01 $^{+ 0.03 }_{- 0.02 }$ & 0.39  $\pm$  0.03 $^{+ 0.08 }_{- 0.06 }$ \\[3pt] 
0.30 & 0.29  $\pm$  0.01 $^{+ 0.02 }_{- 0.02 }$ & 0.31  $\pm$  0.02 $^{+ 0.04 }_{- 0.03 }$ \\[3pt] 
0.34 & 0.21  $\pm$  0.01 $^{+ 0.02 }_{- 0.02 }$ & 0.24  $\pm$  0.01 $^{+ 0.03 }_{- 0.04 }$ \\[3pt] 
0.38 & 0.14  $\pm$  0.00 $^{+ 0.02 }_{- 0.02 }$ & 0.15  $\pm$  0.01 $^{+ 0.02 }_{- 0.02 }$ \\[3pt] 
\hline
\end{tabular}
\end{center}
\end{table}

\newpage

\begin{figure}[!h]
\vspace{0.8cm}
\includegraphics[width=8.5cm,height=10.5cm]{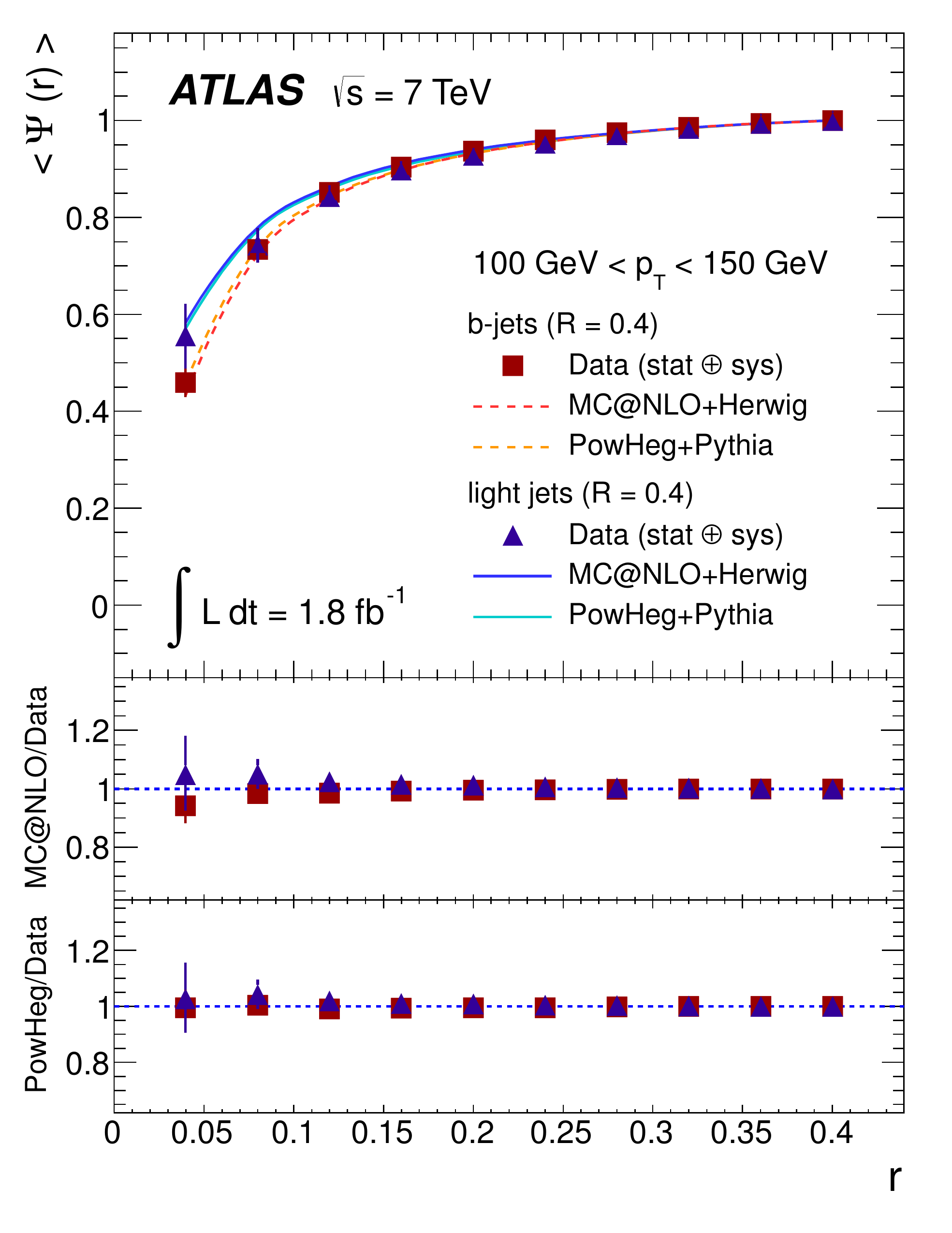}
\caption{Integrated jet shapes $\langle\Psi(r)\rangle$ as a function of the radius $r$ for light jets (triangles) and $b$-jets (squares). The data are compared to \textsc{MC@NLO+Herwig} and \textsc{Powheg+Pythia} event generators for $100 \GeV < p_{\mathrm{T}} < 150 \GeV$. The uncertainties shown include statistical and systematic sources, added in quadrature.}
\label{fig:int5}
\vspace{-0.15cm}
\end{figure}

\begin{table}[!h]
\vspace{0.95cm}
\caption{Unfolded values for $\langle\Psi(r)\rangle$, together with statistical and systematic uncertainties for $100 \GeV < p_{\mathrm{T}} < 150 \GeV$.}
\normalsize
\label{tabInt5}
\begin{center}
\begin{tabular}{ccc}
\hline
$r$ & $\langle\Psi_b (r)\rangle$ [$b$-jets] & $\langle\Psi_{\mathrm{l}} (r)\rangle$ [light jets]\\
\hline
0.04 & 0.459  $\pm$  0.008 $^{+ 0.028 }_{- 0.030 }$ & 0.556  $\pm$  0.022 $^{+ 0.062 }_{- 0.067 }$ \\[3pt] 
0.08 & 0.734  $\pm$  0.005 $^{+ 0.019 }_{- 0.020 }$ & 0.743  $\pm$  0.014 $^{+ 0.033 }_{- 0.036 }$ \\[3pt] 
0.12 & 0.852  $\pm$  0.004 $^{+ 0.013 }_{- 0.012 }$ & 0.843  $\pm$  0.010 $^{+ 0.021 }_{- 0.017 }$ \\[3pt] 
0.16 & 0.904  $\pm$  0.002 $^{+ 0.010 }_{- 0.010 }$ & 0.898  $\pm$  0.007 $^{+ 0.017 }_{- 0.014 }$ \\[3pt] 
0.20 & 0.937  $\pm$  0.002 $^{+ 0.008 }_{- 0.008 }$ & 0.928  $\pm$  0.005 $^{+ 0.014 }_{- 0.011 }$ \\[3pt] 
0.24 & 0.960  $\pm$  0.001 $^{+ 0.006 }_{- 0.006 }$ & 0.954  $\pm$  0.003 $^{+ 0.008 }_{- 0.007 }$ \\[3pt] 
0.28 & 0.975  $\pm$  0.001 $^{+ 0.005 }_{- 0.005 }$ & 0.970  $\pm$  0.002 $^{+ 0.006 }_{- 0.006 }$ \\[3pt] 
0.32 & 0.986  $\pm$  0.001 $^{+ 0.003 }_{- 0.003 }$ & 0.983  $\pm$  0.001 $^{+ 0.003 }_{- 0.003 }$ \\[3pt] 
0.36 & 0.994  $\pm$  0.000 $^{+ 0.001 }_{- 0.001 }$ & 0.994  $\pm$  0.001 $^{+ 0.001 }_{- 0.001 }$ \\[3pt] 
0.40 & 1.000 & 1.000 \\[3pt] 
\hline
\end{tabular}
\end{center}
\end{table}

\clearpage

\begin{figure*}
\centering
\vspace{0.5cm}
\includegraphics[width=16.5cm,height=8.0cm]{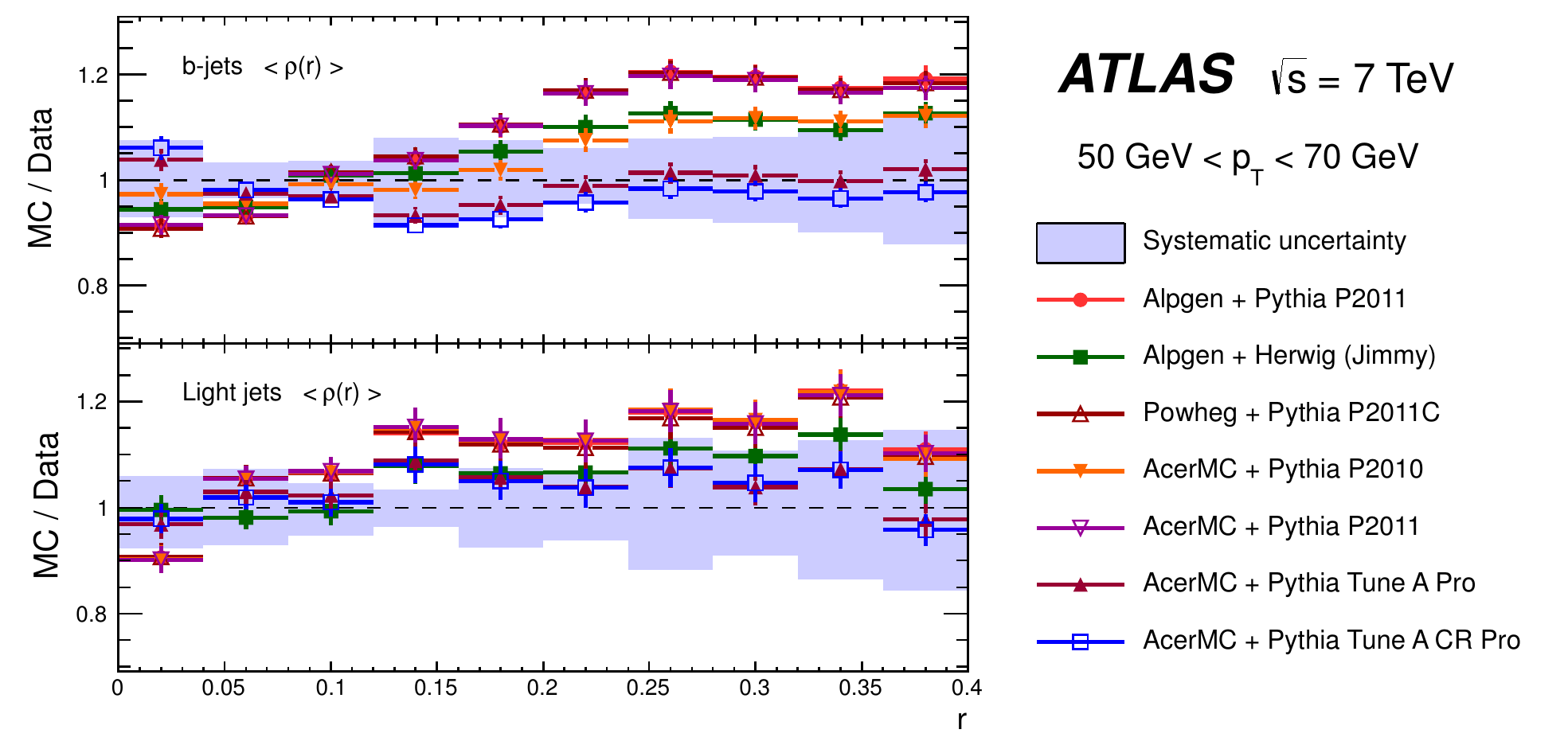}
\caption{Comparison of the $t\bar{t}$ differential jet shape data for $50 \GeV < \pt < 70 \GeV$ with several MC event generators. As stated in the text, \textsc{AcerMC} \cite{acer} coupled to \textsc{Pythia} \cite{pythia} with the \textsc{A Pro} and \textsc{A CR Pro} tunes \cite{tuneA1,tuneA2} give the best description of the data, while the \textsc{Perugia 2011} \cite{Perugia} tunes are found to be slightly disfavoured. \textsc{Alpgen + Jimmy} \cite{alpgen,Jimmy} provides an intermediate description.}
\label{fig:mcComps}
\end{figure*}

\onecolumn
\clearpage

% ATLAS Collaboration author list
% Data extracted on 12-Nov-2013 for paper reference STDM-2011-48
\begin{flushleft}
{\Large The ATLAS Collaboration}

\bigskip

G.~Aad$^{\rm 48}$,
T.~Abajyan$^{\rm 21}$,
B.~Abbott$^{\rm 112}$,
J.~Abdallah$^{\rm 12}$,
S.~Abdel~Khalek$^{\rm 116}$,
A.A.~Abdelalim$^{\rm 49}$,
O.~Abdinov$^{\rm 11}$,
R.~Aben$^{\rm 106}$,
B.~Abi$^{\rm 113}$,
M.~Abolins$^{\rm 89}$
% ...
\bigskip
\\
% ...
$^{11}$ Institute of Physics, Azerbaijan Academy of Sciences, Baku, Azerbaijan\\
$^{12}$ Institut de F{\'\i}sica d'Altes Energies and Departament de F{\'\i}sica de la Universitat Aut{\`o}noma de Barcelona, Barcelona, Spain\\
% ...
$^{21}$ Physikalisches Institut, University of Bonn, Bonn, Germany\\
% ...
$^{48}$ Fakult{\"a}t f{\"u}r Mathematik und Physik, Albert-Ludwigs-Universit{\"a}t, Freiburg, Germany\\
$^{49}$ Section de Physique, Universit{\'e} de Gen{\`e}ve, Geneva, Switzerland\\
% ...
$^{89}$ Department of Physics and Astronomy, Michigan State University, East Lansing MI, United States of America\\
% ...
$^{106}$ Nikhef National Institute for Subatomic Physics and University of Amsterdam, Amsterdam, Netherlands\\
% ...
$^{112}$ Homer L. Dodge Department of Physics and Astronomy, University of Oklahoma, Norman OK, United States of America\\
$^{113}$ Department of Physics, Oklahoma State University, Stillwater OK, United States of America\\
% ...
$^{116}$ LAL, Universit{\'e} Paris-Sud and CNRS/IN2P3, Orsay, France\\
% ...
$^{*}$ Deceased
\end{flushleft}


\clearpage


\begin{thebibliography}{}
\bibitem{cone} J. Chay, S. D. Ellis, Phys.Rev. D \textbf{55} 2728-2735 (1997). arXiv:9607464 [hep-ph]
\bibitem{kt} S. Catani, Y. L. Dokshitzer, M. H. Seymour and B. R. Webber, Nucl. Phys. B \textbf{406} 187 (1993). http://cds.cern.ch/search?sysno=000162968CER
\bibitem{ellis} S.D. Ellis, Z. Kunszt and D. Soper, Phys. Rev. Lett. \textbf{69}, 3615 (1992). arXiv:9208249 [hep-ph]
\bibitem{lep} OPAL Collaboration, R. Akers \emph{et al.} Z. Phys. C \textbf{63}, 197 (1994). http://cds.cern.ch/search?sysno=000178684CER
\bibitem{d0} D0 Collaboration, S. Abachi \emph{et al.} Phys. Lett. B \textbf{357}, 500 (1995). 
\bibitem{h1} H1 Collaboration, C. Adloff \emph{et al.} Nucl. Phys. B \textbf{545}, 3 (1999). arXiv:9901010 [hep-ex]
\bibitem{hera2} ZEUS Collaboration, J. Breitweg \emph{et al.} Eur. Phys. J. C \textbf{8}, 367 (1999). arXiv:9804001 [hep-ex]
\bibitem{chek} ZEUS Collaboration, S. Chekanov \emph{et al.}, Nucl. Phys. B \textbf{700}, 3 (2004). arXiv:hep-ex/0405065
\bibitem{cdf01} CDF Collaboration, D. Acosta \emph{et al.} Phys. Rev. D \textbf{71}, 112002 (2005). arXiv:0505013 [hep-ex]
\bibitem{lhc}  ATLAS Collaboration, Phys. Rev. D \textbf{83}, 052003 (2011). arXiv:1101.0070 [hep-ex]
\bibitem{cms} CMS Collaboration,  J. High Energy Phys. \textbf{06}, 160 (2012). arXiv:1204.3170 [hep-ex]
\bibitem{kellis} R. K. Ellis, W. J. Stirling and B. R. Webber. QCD and Collider Physics, Cambridge University Press.
\bibitem{leadLog} W.T. Giele, E.W.N. Glover, D.A. Kosower, Phys. Rev. D \textbf{57} 1878 (1998). arXiv:9706210 [hep-ph]
\bibitem{boost1} J.M. Butterworth, A.R. Davison, M. Rubin and G.P. Salam, Phys. Rev. Lett. \textbf{100}, 242001 (2008). arXiv:0802.2470 [hep-ph]
\bibitem{boost2} D. Kaplan \emph{et al.}, Phys. Rev. Lett. \textbf{101}, 142001 (2008). arXiv:0806.0848 [hep-ph]
\bibitem{boost3} G. Soyez, G.P. Salam, J. Kim, S. Dutta and M. Cacciari, CERN-PH-TH/2012-300. arXiv:1211.2811 [hep-ph]
\bibitem{gluinos} ATLAS Collaboration, J. High Energy Phys. \textbf{12}, 086 (2012). arXiv:1210.4813 [hep-ex]
\bibitem{CDF} CDF Collaboration, A. Abulencia \emph{et al.} Phys. Rev. D \textbf{78}, 072005 (2008). arXiv:0806.1699 [hep-ex]
\bibitem{detector} ATLAS Collaboration, J. Instrum. \textbf{3}, S08003 (2008). http://iopscience.iop.org/1748-0221/3/08/S08003
\bibitem{atlasTrigger} ATLAS Collaboration, Eur. Phys. J. C \textbf{72} 1849 (2012). arXiv:1110.1530 [hep-ex]
\bibitem{geant1} GEANT4 Collaboration, S. Agostinelli \emph{et al.} Nucl. Instrum. Methods A \textbf{506}, 250-303 (2003). http://cds.cern.ch/search?sysno=002361110CER
\bibitem{geant2} ATLAS Collaboration, Eur. Phys. J. C \textbf{70}, 823 (2010). arXiv:1005.4568 [physics.ins-det]
\bibitem{mcnlo} S. Frixione \emph{et al.}, J. High Energy Phys. \textbf{01}, 053 (2011). arXiv:1010.0568 [hep-ph]
\bibitem{powheg} S. Frixione, P. Nason and C. Oleari, J. High Energy Phys. \textbf{0711}, 070 (2007). arXiv:0709.2092 [hep-ph]
\bibitem{Herwig} M. B\"ahr \emph{et al.}, Eur. Phys. J. C\textbf{58}, 639 (2008). arXiv:0803.0883 [hep-ph]
\bibitem{cluster} R. D. Field, S. Wolfram, Nucl. Phys. B \textbf{213} 65 (1983). 
\bibitem{pdf1} J. Pumplin \emph{et al.}, J. High Energy Phys. \textbf{0207}, 012 (2002). arXiv:0201195 [hep-ph]
\bibitem{Jimmy} J. Butterworth, J. Forshaw and M. Seymour, Z. Phys. C \textbf{72}, 637 (1996). arXiv:9601371 [hep-ph]
\bibitem{auet1} ATLAS Collaboration, ATL-PHYS-PUB-2010-014. http://cds.cern.ch/record/1303025
\bibitem{xsec1l} ATLAS Collaboration, Phys. Lett. B \textbf{711}, 244 (2012). arXiv:1201.1889 [hep-ex]
\bibitem{xsec2l} ATLAS Collaboration, J. High Energy Phys. \textbf{1205} 059 (2012). arXiv:1202.4892 [hep-ex]
\bibitem{spinCorr} ATLAS Collaboration, Phys. Rev. Lett. \textbf{108}, 212001 (2012). arXiv:1203.4081 [hep-ex]
\bibitem{pythia} T. Sj\"ostrand \emph{et al.}, Comput. Phys. Commun. \textbf{135}, 238 (2001). arXiv:0010017 [hep-ph]
\bibitem{pdf2} A. Sherstnev and R.S. Thorne, Eur. Phys. J. C \textbf{55}, 553-575 (2008). arXiv:0711.2473 [hep-ph]
\bibitem{acer} B. Kersevan and E. Richter-Was, Comput. Phys. Commun. \textbf{149}, 142-194 (2003). arXiv:0201302 [hep-ph]
\bibitem{Perugia} P. Z. Skands, Phys. Rev. D \textbf{82}, 074018 (2010). arXiv:1005.3457 [hep-ph]
\bibitem{lund} B. Andersson, G. Gustafson, G. Ingelman, T. Sj\"ostrand, Phys. Rep. \textbf{97} 31 (1983).
\bibitem{ambt1} ATLAS Collaboration, New J. Phys. \textbf{13}, 053033 (2011). arXiv:1012.5104 [hep-ex]
\bibitem{alpgen} M.L. Mangano, M. Moretti, F. Piccinini, R. Pittau, A. Polosa, J. High Energy Phys. \textbf{0307}, 001 (2003). arXiv:0206293 [hep-ph]
\bibitem{pdf3} J. Pumplin \emph{et al.}, J. High Energy Phys. \textbf{02}, 032 (2006). arXiv:0512167 [hep-ph]
\bibitem{HATHOR} M. Aliev \emph{et al.}, Comput. Phys. Commun. \textbf{182}, 1034 (2011). arXiv:1007.1327 [hep-ph]
\bibitem{kidonakis1} N. Kidonakis, Phys. Rev. D \textbf{83}, 091503 (2011). arXiv:1103.2792 [hep-ph]
\bibitem{kidonakis2} N. Kidonakis, Phys. Rev. D \textbf{81}, 054028 (2010). arXiv:1001.5034 [hep-ph]
\bibitem{kidonakis3} N. Kidonakis, Phys. Rev. D \textbf{82}, 054018 (2010). arXiv:1005.4451 [hep-ph]
\bibitem{fewz} R. Gavin \emph{et al.} , Comput. Phys. Commun. \textbf{182}, 2388 (2011). arXiv:1011.3540 [hep-ph]
\bibitem{elec} ATLAS Collaboration, Eur. Phys. J. C \textbf{72}, 1909 (2012). arXiv:1110.3174 [hep-ex]
\bibitem{jets} M. Cacciari, G.P. Salam, G. Soyez, J. High Energy Phys. \textbf{063}, 0804 (2008). arXiv:0802.1189 [hep-ph]
\bibitem{fastjet} M. Cacciari, G.P. Salam, G. Soyez, Eur. Phys. J. C \textbf{72}, 1896 (2012). arXiv:1111.6097 [hep-ph]
\bibitem{topmass} ATLAS Collaboration, Eur. Phys. J. C \textbf{72}, 2046 (2012). arXiv:1203.5755 [hep-ex]
\bibitem{atlasJets} ATLAS Collaboration, Eur. Phys. J. C \textbf{71}, 1512 (2011). arXiv:1009.5908 [hep-ex]
\bibitem{lampl} W. Lampl \emph{et al.}, ATLAS-LARG-PUB-2008-002. http://cds.cern.ch/record/1099735
\bibitem{aleksa} M. Aleksa \emph{et al.}, ATL-LARG-PUB-2006-003. http://cds.cern.ch/record/942528
\bibitem{aharouche} M. Aharrouche \emph{et al.}, Eur. Phys. J. C \textbf{70}, 1193 (2010). arXiv:1007.5423 [physics.ins-det]
\bibitem{JES} ATLAS Collaboration, Eur. Phys. J. C \textbf{73}, 2304 (2013). arXiv:1112.6426 [hep-ex]
\bibitem{atlasBxsec} ATLAS Collaboration, Eur. Phys. J. C \textbf{71}, 1846 (2011). arXiv:1109.6833 [hep-ex]
\bibitem{etmiss} ATLAS Collaboration, Eur. Phys. J. C \textbf{72}, 1844 (2012). arXiv:1108.5602 [hep-ex]
\bibitem{lc1} C. Issever, K. Borras and D. Wegener, Nucl. Instrum. Methods A \textbf{545}, 803 (2005). arXiv:0408129 [physics.ins-det] 
\bibitem{lc2} ATLAS Collaboration, ATL-LARG-PUB-2009-001-2. http://cds.cern.ch/record/1112035
\bibitem{menke} ATLAS Collaboration, ATLAS-CONF-2010-053. http://cds.cern.ch/record/1281310
\bibitem{qcdBkg} ATLAS Collaboration, Phys. Lett. B \textbf{717}, 330 (2012). arXiv:1205.3130 [hep-ex]
\bibitem{kalman} R. Fruehwirth, Nucl. Instrum. Methods A \textbf{262}, 444 (1987).
\bibitem{btagAtl} ATLAS Collaboration, ATLAS-CONF-2011-102. http://cds.cern.ch/record/1369219
\bibitem{btagAtl2} ATLAS Collaboration, ATLAS-CONF-2012-043. http://cds.cern.ch/record/1435197
\bibitem{bayes} G. D'Agostini, Nucl. Instrum. Methods A \textbf{362}, 487-498 (1995). 
\bibitem{roounfold} T. Adye, arXiv:1105.1160 (physics.data-an). 
\bibitem{boostedJets} ATLAS Collaboration, Phys. Rev. D \textbf{86}, 072006 (2012). arXiv:1206.5369 [hep-ex]
\bibitem{singleHadron} ATLAS Collaboration, Eur. Phys. J. C \textbf{73}, 2305 (2013). arXiv:1203.1302 [hep-ex]
\bibitem{jetResol} ATLAS Collaboration, Eur. Phys. J. C \textbf{73}, 2306 (2013). arXiv:1210.6210 [hep-ex]
\bibitem{tuneA1} M. Albrow \emph{et al.} (TeV4LHC QCD Working Group). Fermilab-Conf-06-359 (2006). arXiv:0610012 [hep-ph]
\bibitem{tuneA2} R. D. Field, CDF Note 6403. arXiv:0201192 [hep-ph]
\end{thebibliography}
\end{document}